\documentclass{article}

\usepackage{graphicx,amsmath,amssymb,chemarr,color,cite}

\newcommand{\beq}{\begin{equation}}
\newcommand{\eeq}{\end{equation}}
\newcommand{\beqn}{\begin{eqnarray}}
\newcommand{\eeqn}{\end{eqnarray}}
\newcommand{\elabel}[1]{\label{eq:#1}}
\newcommand{\eref}[1]{Eqn.\ \ref{eq:#1}}
\newcommand{\erefs}[2]{Eqns.\ \ref{eq:#1} and \ref{eq:#2}}
\newcommand{\erefn}[2]{Eqns.\ \ref{eq:#1}-\ref{eq:#2}}

\newcommand{\fref}[1]{Fig.\ \ref{fig:#1}}

\newcommand{\slabel}[1]{\label{sec:#1}}
\newcommand{\sref}[1]{Sec.\ \ref{sec:#1}}
\newcommand{\alabel}[1]{\label{app:#1}}
\newcommand{\aref}[1]{Appendix \ref{app:#1}}

\newcommand{\ket}[1]{|#1\rangle}
\newcommand{\bra}[1]{\langle#1|}
\newcommand{\ip}[2]{\langle#1|#2\rangle}
\newcommand{\avg}[1]{\langle#1\rangle}
\newcommand{\stir}[2]{\left\{\begin{matrix}#1\\#2\end{matrix}\right\}}

\renewcommand{\L}{{\cal L}}
\renewcommand{\d}{\mathchar'26\mkern-12mu d}
\newcommand{\ap}{\hat{a}^+}
\newcommand{\am}{\hat{a}^-}
\newcommand{\Lh}{\hat{\cal L}}
\newcommand{\bh}{\hat{\beta}}
\newcommand{\bb}{\bar{\beta}}
\newcommand{\Gh}{\hat{\Gamma}}
\newcommand{\Dh}{\hat{\Delta}}

\newcommand{\mol}[1]{\mathcal{#1}}
\newcommand{\abs}[1]{\left|{#1}\right|}
\newcommand{\I}{I\left[\alpha,m\right]}

\begin{document}

\title{Spatial partitioning improves the reliability of biochemical signaling}
\author{Andrew Mugler$^1$,
	Filipe Tostevin$^{1}$,
	and Pieter Rein ten Wolde$^1$
	\vspace{.15in} \\
\small $^1$FOM Institute AMOLF, Science Park 104, 1098 XG Amsterdam, The Netherlands}
\date{}
\maketitle

\begin{abstract} 
Spatial heterogeneity is a hallmark of living systems, even at the molecular
scale in individual cells. A key example is the partitioning of membrane-bound
proteins via lipid domain formation or cytoskeleton-induced corralling. Yet the
impact of this spatial heterogeneity on biochemical signaling processes is
poorly understood. Here we demonstrate that partitioning improves the
reliability of biochemical signaling. We exactly solve a stochastic model
describing a ubiquitous motif in membrane signaling. The solution reveals that
partitioning improves signaling reliability via two effects: it moderates the
non-linearity of the switching response, and it reduces noise in the response by
suppressing correlations between molecules. An optimal partition size arises
from a trade-off between minimizing the number of proteins per partition to
improve signaling reliability and ensuring sufficient proteins per partition to
maintain signal propagation. The predicted optimal partition size agrees
quantitatively with experimentally observed systems. These results persist in
spatial simulations with explicit diffusion barriers. Our findings suggest that
molecular partitioning is not merely a consequence of the complexity of cellular
substructures, but also plays an important functional role in cell signaling.
\end{abstract}

The cell membrane is a nexus of information processing.  Once regarded
as a simple barrier between a cell and its surroundings, it is now clear that
the membrane is a hotspot of molecular activity, where signals are integrated
and modulated even before being relayed to the inside of the cell
\cite{Grecco11}.  Moreover, the membrane itself is structurally complex.
Regions enriched in glycosphingolipids, cholesterol, and other membrane
components, often called lipid rafts, transiently assemble and float within the
surrounding bilayer \cite{Eggeling09}, providing platforms for molecular
interaction \cite{Lingwood10}.  Additionally, interaction of the membrane with
the underlying actin cytoskeleton forms compartments in which molecules are
transiently trapped \cite{Kusumi96, Kusumi10}.  These membrane sub-domains
create a highly heterogeneous environment in which molecules are far from
well mixed, and it is currently unclear what effect this heterogeneity has on
cell signaling.

Membrane sub-domains are thought to play a dominant role in the observed
aggregation of signaling molecules into clusters \cite{Kholodenko10}.
Interestingly, these clusters have a characteristic size of only a few
molecules.  For example, the GPI-anchored receptor CD59 is observed to form
clusters of three to nine molecules upon interaction with the cytoskeleton and
lipid rafts \cite{Suzuki07a, Suzuki07b}.  Similarly, the well-studied
membrane-bound GTPase Ras forms  clusters of six to eight molecules which also
depend on interactions with the cytoskeleton and rafts \cite{Prior03,
Plowman05}.  Despite the important findings that aggregation of proteins induced
by sub-domains can affect reaction kinetics \cite{Kalay12}, enhance
oligomerization \cite{Grecco11}, modulate downstream responses \cite{Mugler12,
Tian07} and enhance signal fidelity \cite{Tian07, Gurry09}, the origin of this
characteristic size remains unknown.  While it is quite possible that these
domains owe their size to a thermodynamic or structural origin, we here address
the question of whether this size can be optimized for signaling performance. We
find that the partitioning imposed by sub-domains gives rise to a trade-off in
cell signaling, from which an optimal size of a few molecules emerges naturally,
suggesting that reliable signaling is intimately tied to the spatial structure
of the membrane.

We study via stochastic analysis and spatial simulation a model that is directly
motivated by both CD59 and Ras signaling at the membrane.  Stimulated CD59
receptors induce the switching of several Src-family kinases from an
unphosphorylated to a phosphorylated state \cite{Suzuki07a, Suzuki07b}.
Similarly, stimulated EGF receptors
induce the switching of Ras proteins from an inactive GDP-loaded state to an
active GTP-loaded state \cite{Tian07}.
We therefore study the simple and ubiquitous motif of
coupled switching reactions, in which the activation of one species (the
receptor) triggers the activation of a second species (the downstream effector).

We exactly solve this stochastic model of coupled switching reactions, and we
use the solution to compare signaling reliability in a spatially-partitioned
system to that in a well-mixed system. We demonstrate that partitioning can
improve signaling performance by generating a more graded input-output relation
and by reducing the noise in the signaling response. This latter effect comes
about because partitioning reduces the correlations between the states of the
different output molecules. On the other hand, the stochastic exchange of
proteins between partitions can generate configurations which isolate molecules
and exclude them from the signaling process, thereby reducing the dynamic range
of the response and increasing the output noise. The trade-off between these two
effects results in an optimal partition size that agrees well with cluster
sizes of signaling proteins that are observed experimentally \cite{Suzuki07a,
Suzuki07b,Prior03, Plowman05}, suggesting that cluster sizes are tuned so as to
maximize information transmission.

\section{Results}

\begin{figure}[tb]
	\begin{center}
		\includegraphics[width=.75\textwidth]{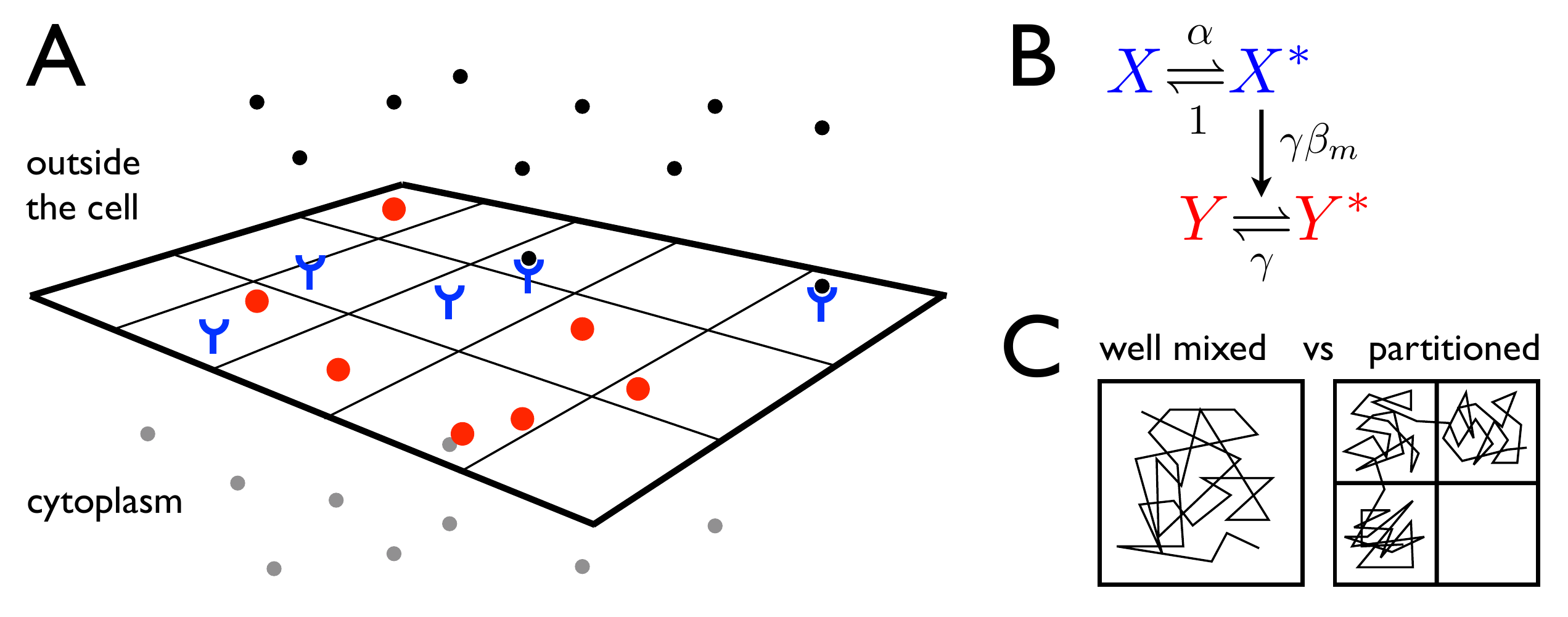}
	\end{center}
	\caption{	\label{fig:fig1}
		Schematic depiction of the model system. 
		{\bf A} We consider a model representative
		of signal detection by receptors 
		and signal transmission at the cell membrane. 
		{\bf B} The model consists of two molecular species ($\mol{X}$ and
		$\mol{Y}$) which can each exist in active ($X^*$, $Y^*$)
		or inactive ($X$,
		$Y$) states. Molecules in the $X$ state are activated by
		the external signal
		of strength $\alpha$, and active $X^*$ molecules
		subsequently activate $Y$
		molecules.  
		{\bf C} We consider these reactions taking place
		in a single domain with all
		components well mixed, or in a domain consisting of smaller
		compartments
		which are each individually well mixed but between
		which no interaction is
		possible. The total system volumes in the two scenarios are equal and
		assumed to scale with the number of $\mol{X}$ molecules.
	}
\end{figure}

We model two coupled molecular species at the membrane, as depicted in Fig.\
\ref{fig:fig1}A. A membrane-bound receptor (e.g.\ CD59 or EGF receptor) is
activated via ligand stimulation, and the active receptor in turn activates a
membrane-bound effector (e.g.\ a Src-family kinase or Ras). A reaction scheme
representing these processes is shown in Fig.\ \ref{fig:fig1}B, and consists of
two protein species: the receptor $\mol{X}$ and the downstream effector
$\mol{Y}$. The switching of $\mol{X}$ molecules from the $X$ to the $X^*$ state
is driven by an external signal of strength $\alpha$.  Active $X^*$ molecules
act on inactive $Y$ molecules and promote switching to the $Y^*$ state.
Deactivation of both active protein species occurs spontaneously and
independently.

We will be concerned with how the network response, the number of active $Y^*$
molecules as a function of the input signal $\alpha$, is affected by the spatial
structure of the system.  In particular we ask how partitioning of the reaction
system into non-interacting sub-domains affects the reliability of signal
transmission, which is determined by two principal factors: the input-output
response and the output noise; together these properties determine to what
extent different input signals can be reliably resolved from the network
response. We focus on two system configurations, shown in Fig.\ \ref{fig:fig1}C.
In the first case we assume that all molecules are present in a single
well-mixed reaction compartment. In the second case, we consider a system
partitioned into $\pi$ compartments between which no interactions are permitted;
here we take the output of the system to be the total number of active $Y^*$
molecules in all compartments.
This choice of output corresponds to a readout of
the $Y^*$ signal by, e.g., a cytosolic component whose diffusion is much faster
than the diffusion and signaling of $\cal{X}$ and $\cal{Y}$ on the membrane.
In the partitioned system, we will for simplicity first assume that the
molecules are uniformly and statically distributed among compartments. However,
recognizing that this scenario will not generally be realized inside cells,
we will later relax this assumption and consider exchange of molecules among
partitions.

We model the dynamics of the well-mixed system, as well as each
compartment within the partitioned system,
using a stochastic equation of the same form.
We denote the total numbers of $\mol{X}$ and $\mol{Y}$ molecules by $M$ and $N$,
respectively, and the numbers of active $X^*$ and $Y^*$ molecules by $m$ and
$n$, respectively. To parameterize the system, we scale units of time by the
deactivation rate of $X^*$, such that the effective deactivation rate is $1$.
Then $\alpha$ denotes the rescaled activation rate of $X$; $\gamma$ is the rate
of deactivation of $Y^*$ relative to that of $X^*$; and $\gamma\beta_m$ is the
activation rate of a given $Y$ molecule for a particular concentration of $X^*$
molecules. The parameter $\alpha$ incorporates the effective strength of the
input signal and determines the mean $X^*$ activity via the occupancy $q \equiv
\avg{m}/M = \alpha/(\alpha+1)$. The precise $m$-dependence of the coupling
function $\beta_m$ will depend on the exact nature of the interactions between
$X^*$ and $Y$ molecules. We take $\beta_m \propto m/v$, with $v$ the volume of
the compartment in which the reactions are taking place. However, our
conclusions are unaffected if we instead take a Michaelis-Menten form $\beta_m
\propto m/(m+vK)$ (\aref{fig}: \fref{mm}).  The total system volume $V$ is
assumed to scale with the total number of $\cal{X}$ molecules, such that $M/V$
is constant. The coupling function in partition $i\in\{1,\dots,\pi\}$ is then
determined by $m_i$, the number of $X^*$ molecules in partition $i$, according
to $\beta^{(i)}_{m}\propto m_i/(V/\pi)=\beta\pi m_i/M$ for constant $\beta$.

The probability of having $m$ proteins in the $X^*$ state and $n$ proteins in
the $Y^*$ state evolves according to the chemical master equation (CME),  
\begin{equation} \label{eq:cme}
	\dot{p}_{mn}=-\left[\mol{L}_m(\alpha,M)
		+\gamma\mol{L}_n(\beta_m,N)\right] p_{mn}, 
\end{equation}
subject to suitable boundary conditions. The nature of the particular set of
reactions in our model (Fig.~\ref{fig:fig1}B) means that the operators
$\mol{L}_m$ and $\mol{L}_n$ have the same form,
\begin{equation}
	\mol{L}_m(\alpha,M)=\alpha\left[1-{\mathbb E}_m^{-1}\right](M-m)
		+\left[1-{\mathbb E}_m^{+1}\right]m, 
\end{equation}
where ${\mathbb E}_m^{i}f(m)=f(m+i)$ defines the step operator. Despite the
appearance of terms containing the product $mn$ in the operator
$\mol{L}_n(\beta_m,N)$, which make the direct calculation of moments of
$p_{mn}$
from the CME impossible, an exact solution to (\ref{eq:cme}) can be found for
arbitrary $\beta_m$ using the method of spectral expansion
\cite{Walczak09,Mugler09} as described in \aref{spec}.

\subsection{Partitioning leads to a more graded response}

We begin by analyzing the behavior of a minimal system with $M=N=2$. In the
well-mixed system, all molecules are contained within $\pi=1$ domain of volume
$V$. In the partitioned system, $\pi=2$ subdomains with volume $V/2$ each
contain one $\mol{X}$ and one $\mol{Y}$ molecule.

\begin{figure}[tb]
	\begin{center}
		\includegraphics[width=0.75\textwidth]{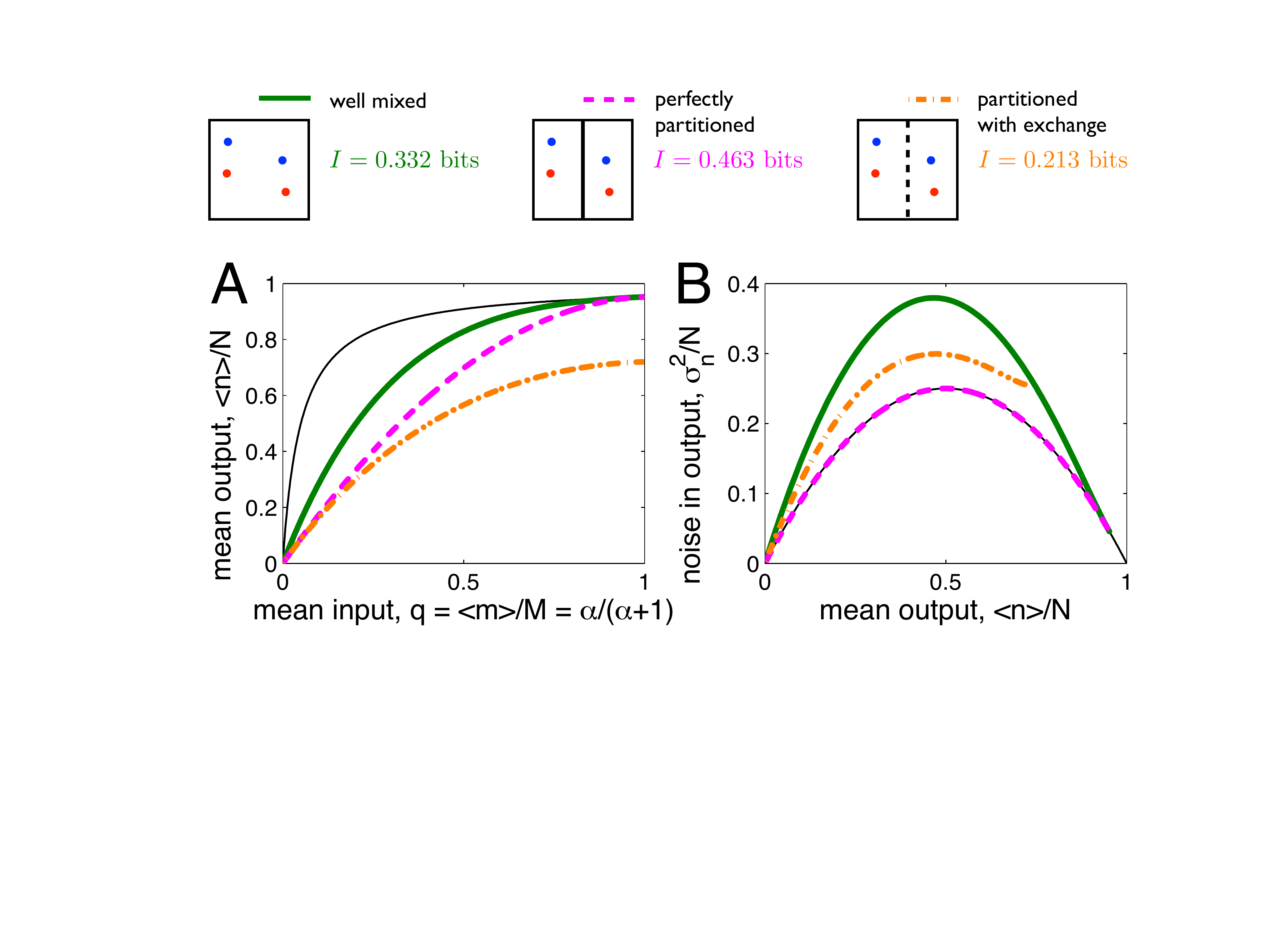}
	\end{center}
	\caption{	\label{fig:fig2}
		Spatial partitioning improves signaling performance. 
		{\bf A} The mean response $\avg{n}/N$ as a function of the mean $X^*$
		activity $q=\avg{m}/M=\alpha/(\alpha+1)$, and {\bf B}
		the output variance
		$\sigma^2_n$ as a function of the mean response, plotted for a
		well-mixed system with $M=N=2$ (thick solid) and a
		partitioned system of
		$\pi=2$ compartments, each containing one $\mol{X}$ and one $\mol{Y}$
		molecule (thick dashed). Partitioning linearizes the output
		response and
		reduces noise across the full range of responses, leading to a higher
		transmitted information.  The thin solid curves show the mean
		field response
		$\avg{n}/N=\beta q/(\beta q+1)$ in A and the binomial noise limit
		(\ref{eq:variance_partitioned}) in B.
		Allowing exchange of molecules between compartments (thick dot-dashed)
		compresses the output response and increases the noise compared to the
		perfectly partitioned system,
		dramatically reducing information transmission.
		Here $\beta=20$ and $\gamma=1$.
	}
\end{figure}

We first focus on the mean response $\avg{n}$. In the limits of small or large
$\alpha$ the mean response is the same in both the partitioned and mixed
systems, $\avg{n}/N\to0$ and $\avg{n}/N\to\beta/(\beta+1)$ respectively.
However, at all intermediate values of $\alpha$, the mean response of the
well-mixed system is larger than that of the partitioned system; equivalently,
the partitioned system exhibits a more graded response than the well-mixed
system to changes in the input signal (see Fig.\ \ref{fig:fig2}A, thick solid
and dashed curves).  The more graded response is due to higher fluctuations in
$X^*$ activity.  When $\alpha\to0$ or $\alpha\to\infty$, all ${\cal X}$
molecules are inactive or active, respectively; however at intermediate values
of $\alpha$, the number of active $X^*$ molecules fluctuates.  Partitioning
reduces the number of $\mol{X}$ molecules per reaction compartment, increasing
the relative size of these fluctuations according to
$\sigma^2_m/(M/\pi)^2=\pi q(1-q)/M$. These fluctuations are passed through
the concave dependence of $n$ on $m$, resulting in a smaller mean (via Jensen's
inequality \cite{Jensen06}), and therefore a more linear response curve (see
also \aref{fig}: \fref{linearization}A).

A more graded input-output relation can potentially enhance signaling by
expanding the range of input signals which the network is able to transmit
without saturating the response. However, in order to determine whether this
larger input range can be resolved in the network it is crucial to examine how
the noise in the response is affected.

\subsection{Partitioning reduces noise}

Figure \ref{fig:fig2}B shows the variance of the output $\sigma^2_n$ as a
function of the mean response $\avg{n}$ for the system with $M=N=2$, as the
input signal strength $\alpha$ is varied.  We see that the output noise is
reduced in the partitioned system relative to the well-mixed system across the
full range of response levels.  The noise reduction is surprising: one might
expect that the increased fluctuations in $X^*$ activity that come with
partitioning would propagate to fluctuations in $Y^*$ activity.  Indeed, this is
the case: in a single compartment, as the number of $\mol{X}$ molecules
is reduced, the noise in the output increases
(\aref{fig}: \fref{linearization}B).
However, this effect is overcome by a second effect: partitioning reduces
correlations among output molecules.

\begin{figure}[tb]
	\begin{center}
		\includegraphics[width=0.75\textwidth]{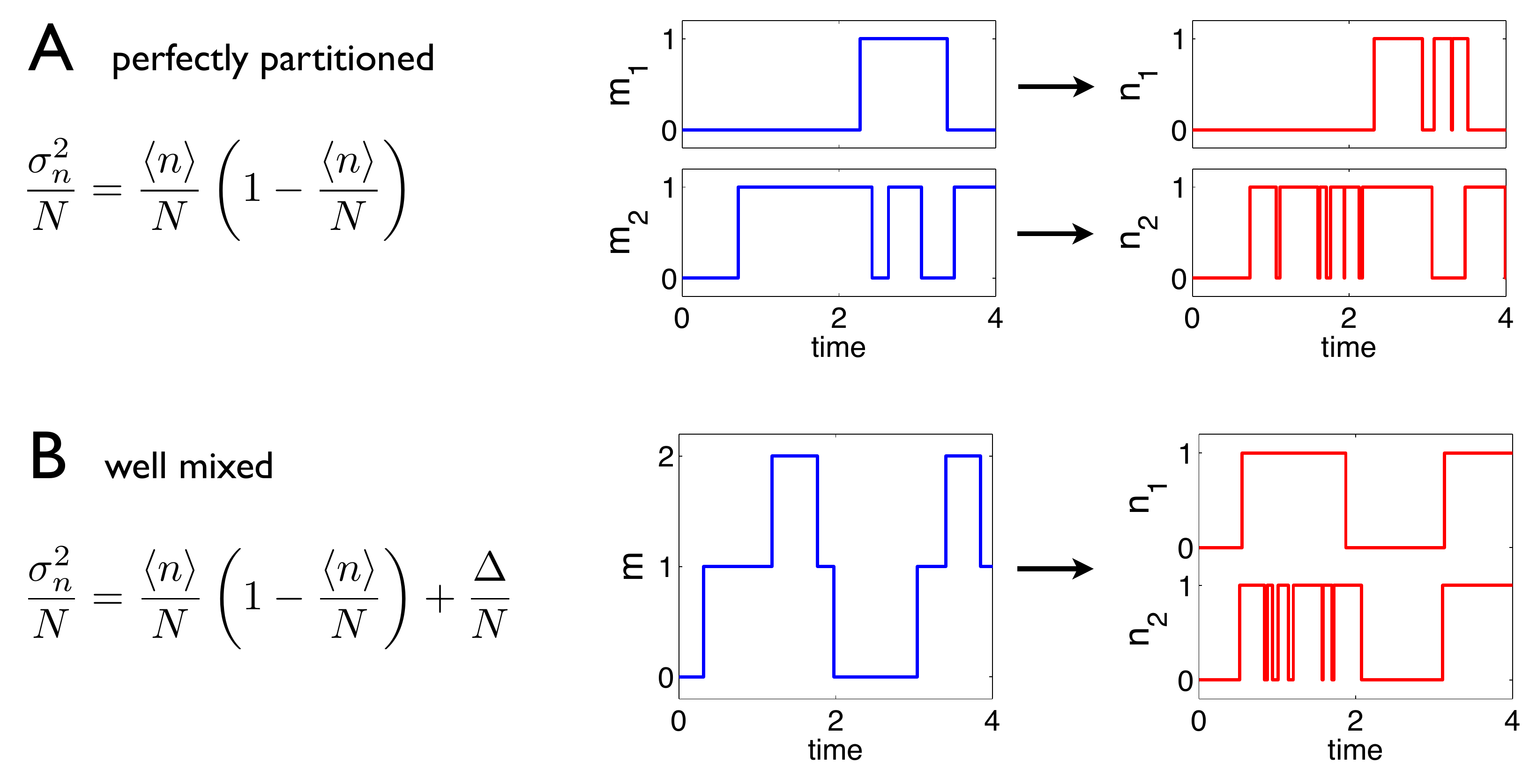}
	\end{center}
	\caption{ \label{fig:fig3}
		Partitioning reduces correlations between output modules. 
		{\bf A} In the partitioned system,
		each $\mol{Y}$ molecule receives an independent
		signal $m_i(t)$. The variance is simply that of independent two-state
		switches. 
		{\bf B} In the well-mixed system,
		each $\mol{Y}$ molecule reacts to the same $m(t)$,
		which leads to correlations between in the states
		of different $\mol{Y}$
		molecules and an increase in the variance $\sigma^2_n$.
		Sample trajectories are generated using parameters as in
		Fig.\ \ref{fig:fig2},
		with $\alpha=1$.
	}
\end{figure}

To see the effect of partitioning on correlations, we  consider the expressions
for the variance.  In the partitioned case, since the two $\mol{Y}$ molecules
switch independently, the variance of $n$ is simply that of a pair of
independent binomial switches with activation probability $\avg{n}/N$,
\begin{equation} \label{eq:variance_partitioned}
	\frac{\sigma^2_n}{N}=\frac{\avg{n}}{N}\left(1-\frac{\avg{n}}{N}\right).
\end{equation}
In contrast, in the well-mixed case the two $\mol{Y}$ molecules are not
independent. Since both are driven by the same set of $\mol{X}$ molecules,
fluctuations in $\beta_m$ lead to correlations between the states of the
two $\mol{Y}$ molecules as their switching becomes more synchronized (see Fig.\
\ref{fig:fig3}). This in turn leads to an increase in the variance, which can be
written as 
\begin{equation} \label{eq:variance_mixed}
	\frac{\sigma^2_n}{N}=\frac{\avg{n}}{N}\left(1-\frac{\avg{n}}{N}\right)
		+\frac{\Delta}{N},
\end{equation}
where $\Delta$ is a correction term accounting for the correlation between
$\mol{Y}$ molecules, which are due to ``extrinsic'' fluctuations in the input
$m(t)$. The functional form of $\Delta$ for any $M$ and $N$ follows directly
from the spectral solution of the CME
(\aref{spec_detail}: \eref{noise_supp}); for $M=N=2$ one finds by inspection that
$\Delta$ is manifestly positive, meaning that correlations increase the noise
across all values of the mean. Importantly, this effect is independent of the
parameters of the switching reactions.

The reduction of noise upon partitioning extends beyond the case of one
$\mol{Y}$ molecule per partition. Indeed the same phenomenon is observed if we
consider larger molecule numbers $M>\pi$ and $N>\pi$, and compare the
well-mixed system to a system with uniform partitioning of the $\mol{X}$ and
$\mol{Y}$ molecules into the $\pi$ compartments. In the well-mixed case all
$\mol{Y}$ molecules respond to the same signal $m(t)$, and hence are correlated
with all other $\mol{Y}$ molecules in the system. By contrast, in the
partitioned case the $N/\pi>1$ $\cal{Y}$ molecules {\em within} each partition
are correlated, and indeed since the fluctuations in $m_i(t)$ will be larger
than $m(t)$ for the mixed system, such correlations will be stronger; yet, the
$\mol{Y}$ molecules in {\em different} partitions are uncorrelated.  This latter
effect is sufficient to overcome the increase in correlations within each
partition, such that the total noise is reduced.

To see the noise reduction explicitly, we again consider the expression for the
variance. Since the dynamics of different partitions is independent, assuming
that both $M$ and $N$ are multiples of $\pi$, the variance can be written as
\begin{equation} \label{eq:general_variance_partitioned}
	\frac{\sigma^2_n}{N}=\frac{\avg{n}}{N}\left(1-\frac{\avg{n}}{N}\right)
		+\pi\frac{\Delta(\tilde{M},\tilde{N})}{N},
\end{equation}
where $\tilde{M}\equiv M/\pi$ and $\tilde{N}\equiv N/\pi$ are the numbers of
$\cal{X}$ and $\cal{Y}$ molecules per compartment, respectively.  Here, as
before, $\Delta(\tilde{M},\tilde{N})$ represents the additional fluctuations due
to correlations between the states of $\mol{Y}$ molecules within each
compartment.  The $N$-dependence of $\Delta(\tilde{M},\tilde{N})$, which
reflects the number of correlated pairs of $\mol{Y}$ molecules, can be
straightforwardly factored out as
$\Delta(\tilde{M},\tilde{N})=\tilde{N}(\tilde{N}-1)\tilde{\Delta}(\tilde{M})$,
where $\tilde{\Delta}(\tilde{M})$ describes how strongly correlated are
$\mol{Y}$ molecules within each compartment.  The exact form for
$\tilde{\Delta}(\tilde{M})$, while straightforward to calculate for a given
$\tilde{M}$, is difficult to generalize for all $\tilde{M}$; nonetheless,
inspection of numerical and analytic results for specific combinations of
$\tilde{M}$ and $\tilde{N}$ reveals in all cases that increasing $\pi$ leads to
an overall reduction in $\sigma^2_n$.  Additionally, if the switching of
$\mol{Y}$ molecules is much slower than that of $\mol{X}$ molecules,
$\gamma\ll1$, then $\tilde{\Delta}(\tilde{M})$ takes the form
\begin{equation} \label{eq:delta_slow_limit}
	\tilde{\Delta}(\tilde{M})\approx
		\frac{\alpha\beta^2\gamma}{\tilde{M}(1+\alpha+\alpha\beta)^3}
\end{equation}
Inserting this expression into (\ref{eq:general_variance_partitioned}) with
$\tilde{M}=M/\pi$ and $\tilde{N}=N/\pi$, one can straightforwardly see that the
variance is a decreasing function of $\pi$ for $\pi<N$, indicating that the
noise is reduced as the system is more finely partitioned.

\subsection{Partitioning increases information transmission}

We have seen that partitioning has two beneficial effects on signal propagation:
the input-output response becomes more graded, and the output noise at a given
response level is reduced. Together, these effects mean that a larger number of
distinct input signals can be encoded in the network response. To quantify the
ability of the network to transmit signals we calculate the mutual information
$\I$ \cite{Shannon48} between the input and the number of active $Y^*$ molecules, as described in
\aref{mi}. We find that indeed, in the case of $M=N=2$ (Fig.\
\ref{fig:fig2}), $\I$ is significantly larger for the partitioned system
($I=0.463$ bits) than for the well-mixed system ($I=0.332$ bits), confirming
that signal transmission is dramatically improved by partitioning.

\subsection{Exchange between partitions compromises signaling reliability}

\begin{figure}[tb]
	\begin{center}
		\includegraphics[width=0.75\textwidth]{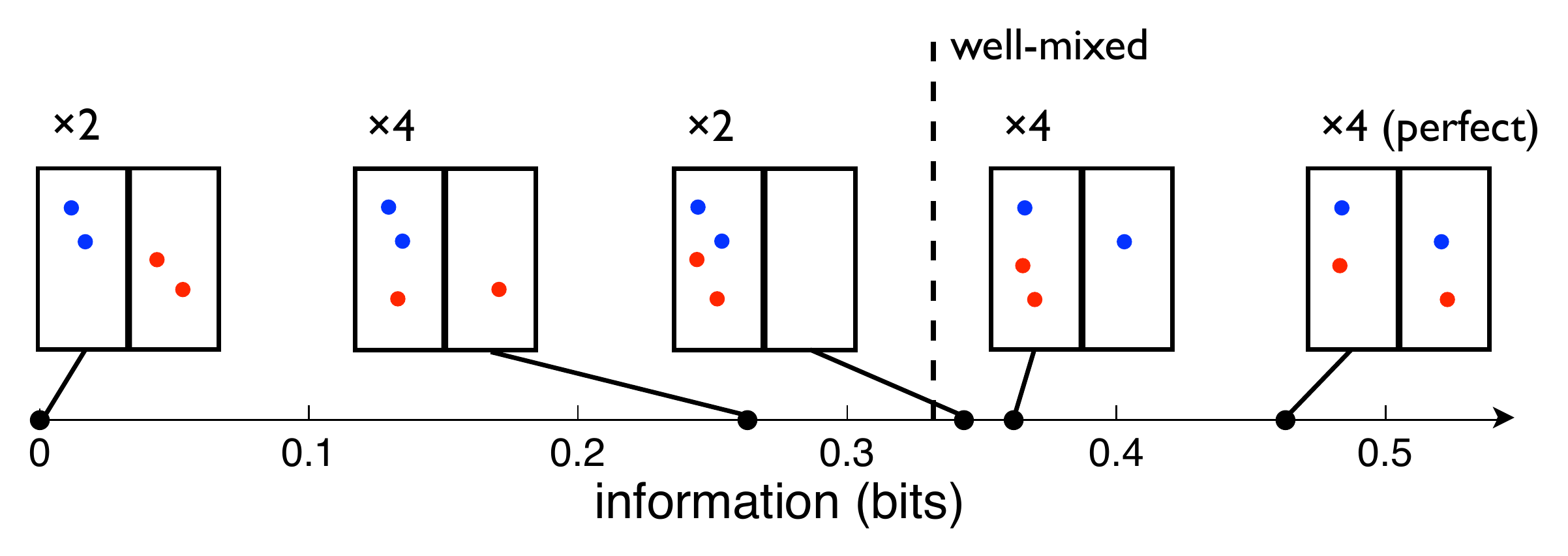}
	\end{center}
	\caption{	\label{fig:fig4}
		Exchange between partitions leads to different configurations of the
		system with a range of signaling performance.
		Multiplicities listed above each configuration are due to symmetry.
		Parameters are as in
		Fig.\ \ref{fig:fig2}.
	}
\end{figure}

Thus far we have considered only the perfectly uniform and stationary
partitioning of molecules. In reality, physical transport processes such as
diffusion will also give rise to a variety of configurations with different
numbers of proteins in each compartment, as depicted in Fig.\ \ref{fig:fig4}.
Each of these configurations will have different properties for the transmission
of the signal from $\alpha$ to $n$. It is therefore important to consider
whether the benefits of partitioning described above persist once these
additional configurations are taken into account.

Single-molecule tracking experiments have revealed that the timescale of
diffusive mixing within a compartment ($\sim$$100$ $\mu$s) is two orders of
magnitude faster than the timescale
of molecular exchange between compartments ($\sim$$10$ ms) \cite{Murase04}.
This observation allows us to treat each configuration as static on the
timescale of mixing, then compute the total response by averaging over all
configurations.
Inherent in this treatment is the assumption that the timescale
of signaling is also faster than that
of exchange between compartments.   We later relax this assumption
using spatially resolved simulations and nonetheless find similar results.

The total response is computed by first enumerating
the possible configurations of $M$ $\mol{X}$ molecules and $N$ $\mol{Y}$
molecules distributed amongst $\pi$ partitions. For each such configuration $c$
we then solve for the output distribution $p_{n|c}$ and combine these
distributions, weighted by the probability $p_c$ of each configuration occurring
if molecules are randomly assigned to different partitions with uniform and
independent probability, to give the overall response distribution
$p_n=\sum_cp_{n|c}p_c$.

Figure \ref{fig:fig2}B (dot-dashed curve) shows that the exchange of molecules
between compartments increases the noise relative to the perfectly partitioned
system considered previously when $M=N=2$. This is because many of the
alternative configurations generated by exchange lead to significant
correlations between the states of the different $\mol{Y}$ molecules.
Nevertheless, we see that the noise remains lower than that of the well-mixed
system, because of the existence of some configurations in which the $\mol{Y}$
molecules are independent. However, the appearance of alternate configurations
also affects the mean response (Fig.\ \ref{fig:fig2}A); in particular, the
appearance of configurations in which $\mol{X}$ and $\mol{Y}$ molecules do not
occupy the same partitions, and hence no signal can be propagated, means that
the maximal output level is reduced.  Given this simultaneous change in both the
input-output function and the noise, it is not immediately clear whether
signaling reliability is improved relative to the well-mixed system. Computing
the mutual information, we see that the information transmitted by the system
with exchange ($I=0.213$ bits) is significantly lower than that for the
well-mixed system ($I=0.332$ bits), showing that the reduction of the output
range compromises signal transmission to an extent which cannot be overcome by
the corresponding reduction in noise.

The decrease in information transmission upon incorporating molecule exchange in
the system with $M=N=2$ is the result of the appearance of suboptimal protein
configurations, for which signal propagation is compromised (or even
impossible). However, the number and performance of such configurations will in
general depend on the relative values of $M$, $N$ and $\pi$ (which need not
equal $M$ or $N$). While molecule exchange may make partitioning unfavorable in
the extreme case of $M=N=2$, for systems with higher protein numbers it can be
beneficial to partition the system into $\pi>1$ compartments, as we will see
next.

\subsection{An optimal partition size}
\begin{figure}[tb]
	\begin{center}
		\includegraphics[width=0.75\textwidth]{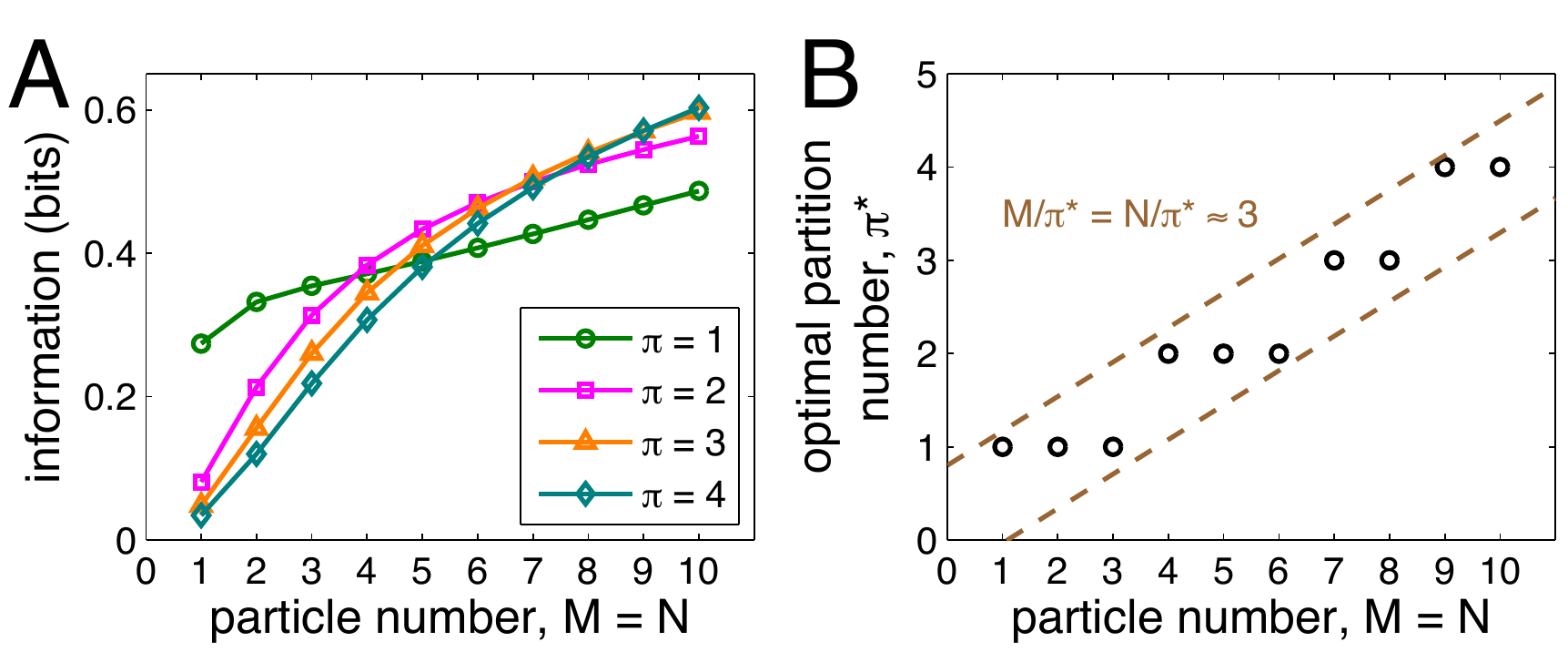}
	\end{center}
	\caption{	\label{fig:fig5}
		An optimal partition size. {\bf A} For
		$M=N>3$ molecules, a system with $\pi>1$
		partitions achieves higher information transmission
		that a well-mixed system
		($\pi=1$). {\bf B} As $M=N$ is increased the optimal partition
		number also increases such that the optimal number of
		proteins per partition $M/\pi^*=N/\pi^*\approx3$ is roughly constant.
		Parameters are as in Fig.\ \ref{fig:fig2}.
	}
\end{figure}

To study the performance of systems with higher protein numbers and different
partition sizes, we compare the information transmission, including molecule
exchange, for different partition numbers $\pi$ as the number of proteins in the
system is varied while holding $M=N$.  Figure~\ref{fig:fig5}A shows that for
$M=N>3$ protein copies, systems with $\pi>1$ partition do indeed outperform the
well-mixed system. Furthermore, as $M=N$ is increased the optimal partition
number also increases such that the optimal number of proteins per partition
$M/\pi^*=N/\pi^*\approx3$ is roughly constant (Fig.\ \ref{fig:fig5}B). This
result is robust to
variations in $\beta$ and $\gamma$: changing each over several orders of
magnitude results in optimal partition sizes in the range
$M/\pi^*=N/\pi^*\sim1$$-$$10$ (\aref{fig}: \fref{betagamma}A and B).  The
assumption of
$M=N$ is also not crucial for this result.  In fact, we find that the value of
$M/\pi^*$ has only a weak dependence on $N$ (\aref{fig}: \fref{piMN}).

The optimal partition size arises from a trade-off between the reliability
and efficiency of signaling. Increasing the number of partitions decreases the
typical number of proteins per partition, which leads to the beneficial effects
of a more graded response and reduced noise, increasing signaling
reliability. On the other hand, due to molecule
exchange, reducing the number of molecules per partition also increases the
probability that any partition contains proteins of only one species that are
therefore excluded from the signaling process, which leads to a reduced
maximal response, reducing signaling efficiency.

The optimal size revealed by our study of $\sim$$1$$-$$10$ molecules per
species per partition shows good quantitative agreement with the observed
aggregation of CD59 receptors ($3$$-$$9$ molecules \cite{Suzuki07a, Suzuki07b})
and Ras proteins ($6$$-$$8$ molecules \cite{Prior03, Plowman05}), which each
signal via the present motif and are known to interact with rafts and the
cytoskelton.  It is of further interest that a recent experiment in which T cell
receptors were artificially partitioned on supported membranes found that the
minimum number of agonist-bound receptors per partition necessary for downstream
signaling is approximately four \cite{Manz11}.

\subsection{An explicitly spatial model}

\begin{figure}[tb]
	\begin{center}
		\includegraphics[width=0.75\textwidth]{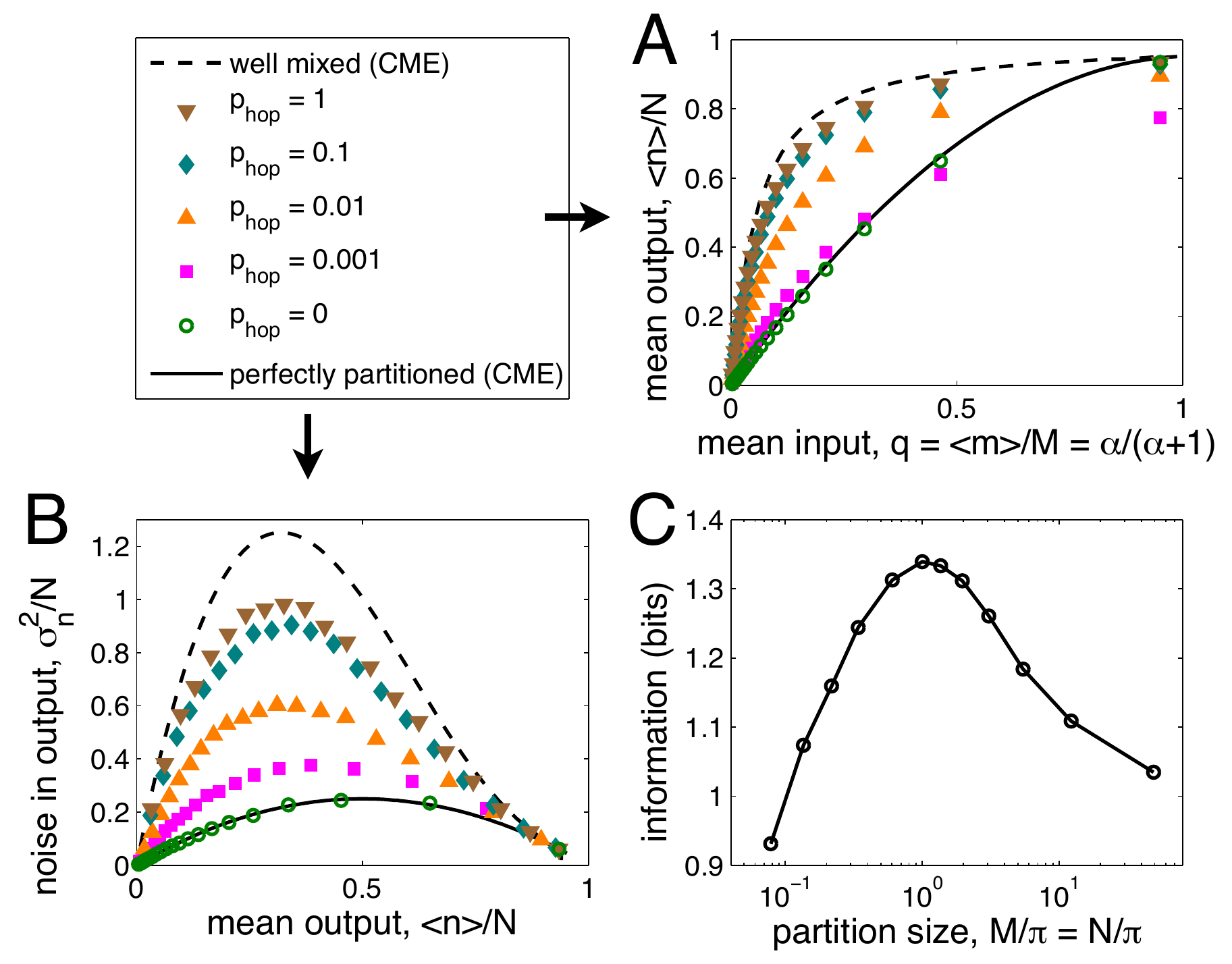}
	\end{center}
	\caption{	\label{fig:fig6}
		The effects of partitioning persist in simulations with explicit
		diffusion.
		As the probability of crossing a diffusion barrier $p_{\rm hop}$ is
		decreased,
		{\bf A} the mean response becomes more graded, and
		{\bf B} the output noise decreases.
		{\bf C} The information transmission has a maximum as a function of
		the partition size.
		Here $M=N=49$, $\beta=20$, $\gamma=1$, the system is $\lambda=70$
		lattice spacings squared, and the ratio of diffusion to reaction
		propensities is $p_D/p_r = 1$.
		In A, $\pi=49$; in B, $p_{\rm hop} = 0.001$, and the partition size
		is varied by taking $\sqrt{\pi}$ from $25$ to $1$.
	}
\end{figure}

Lastly, we confirm that the effects observed in these minimal model systems,
where the contents of each compartment are well-mixed and exchange can occur
between any pair of compartments, persist in a more realistic model in which the
diffusion of molecules in space is included explicitly. We simulate the
diffusion and reaction of $\mol{X}$ and $\mol{Y}$ molecules on a two-dimensional
lattice, as described in \aref{sim}.
The system is partitioned into a number of subdomains by the
introduction of diffusion barriers, which are crossed with a reduced probability
$p_{\rm hop}$ relative to regular diffusion steps on the lattice. Results of
such simulations are shown in Fig.\ \ref{fig:fig6}.

Figure \ref{fig:fig6}A and B reveal that as the strength of the diffusion
barriers is increased,
the mean response becomes more graded, and the variance of $Y^*$ activity is
reduced,
analogous to the two effects observed in the minimal model system
(Fig.\ \ref{fig:fig2}).
When $p_{\rm hop}=0$, one molecule of each species is permanently confined to a
compartment,
producing the graded response predicted for the perfectly partitioned system
(Fig.\ \ref{fig:fig6}A)
and the associated minimal, binomial noise (Fig.\ \ref{fig:fig6}B).  Low but
finite $p_{\rm hop}$ allows exchange of molecules between neighboring
compartments but preserves a separation of timescales between intra- and
inter-compartment mixing.
This results in a graded mean response whose maximal level is reduced
(Fig.\ \ref{fig:fig6}A) and reduced noise (Fig.\ \ref{fig:fig6}B), precisely the
features observed in the minimal model of partitioning with exchange
(Fig.\ \ref{fig:fig2}).  When $p_{\rm hop} = 1$, there are no barriers, and the
system approaches the well-mixed limit (CME).  Interestingly, however, the
response remains more graded and the noise remains lower than the predictions of
the CME due to the finite speed of diffusion (Fig.\ \ref{fig:fig6}A and B), with
agreement only reached when the ratio of diffusion to reaction propensities is
much greater than one.  This observation
reveals that finite diffusion imposes an effective partitioning even when no
actual partitions exist: molecules remain correlated with reaction partners
within a typical distance set by diffusion, but uncorrelated with partners
beyond this distance.  As such, in the context of coupled reversible
modification, we find that slower diffusion can linearize the response and
reduce the noise, thereby improving information
transmission.\footnote{Interestingly, this result is in marked contrast to
the case of boundary establishment in embryonic development, where faster
diffusion reduces noise within each nucleus by washing out bursts of gene
expression in the input signal \cite{Erdmann09}. While in the present system
faster diffusion will similarly reduce any super-Poissonian component of the
noise within each partition individually, this averaging does not reduce the
noise in the total output across all partitions.  In fact, the latter noise is
enhanced with faster diffusion by virtue of increased correlations between
partitions.}
It is important to emphasize, however, that the extent of this effect is much
smaller than for actual partitioning: Fig.\ 6B shows that finite diffusion
reduces the maximal noise by $(1.25-1)/1.25=20\%$, while strong partitioning
($p_{\rm hop}=0.001$) reduces the maximal noise by
$(1.25-0.4)/1.25\approx 70\%$.  Therefore, partitioning, which introduces not
only a slower effective ``hop'' diffusion but also a separation of timescales
between intra- and inter-compartmental mixing, is far more effective at
conveying an information enhancement.

Fig.\ \ref{fig:fig6}C confirms that the transmitted information varies
non-monotonically with the number of barriers in a fixed area, indicating that
an optimal partition size also appears in systems where
space is modeled explicitly.
Like in the minimal model, this optimum persists with changes in $\beta$ and
$\gamma$, spanning the range of $\sim$$1$$-$$10$ molecules per partition (\aref{fig}: \fref{betagamma}C and D).
Fig.\ \ref{fig:fig6}C also provides a measure of the scale of information
transmitted by this motif.  In absolute terms, the optimal information ($1.35$
bits) is consistent with values recently measured for signaling via the
TNF-NF-$\kappa$B pathway ($\sim$$0.5$$-$$1.5$ bits) \cite{Cheong11} and for
patterning in the {\em Drosophila} embryo ($1.5 \pm 0.15$ bits) \cite{Tkacik08}. 
In relative terms, we see that partitioning increases information over the
unpartitioned system by $(1.35-1.04)/1.04 \approx 30\%$ (Fig.\ \ref{fig:fig6}C)
and decreases the maximal noise by $(1-0.4)/1=60\%$ (Fig.\ \ref{fig:fig6}B).
Thus, in both absolute and relative terms, we see that partitioning plays a
critical role in producing informative and reliable membrane signaling.

As a final test, we use simulation to confirm that the effects of
partitioning persist in the presence of features that are more realistic for
signaling systems at the membrane, including extrinsic noise in the input
(\aref{fig}: \fref{ex}) and receptor dimerization
(\aref{fig}: \fref{dimer}).
The fact that the effects of partitioning, including the emergence of an
optimal partition size, are robust to these details further underscores the
generality of our findings.

\section{Discussion}

We have seen that the partitioning of a biochemical signaling system into a
number of non-interacting subsystems improves the reliability of signaling via
two effects. First, the non-linear response of the network means that a
reduction in the number of input molecules translates into a more graded
input-output response. Second, partitioning significantly reduces the noise in
the response by eliminating correlations between the states of the different
output molecules, an effect which, remarkably, overcomes the increase in noise
associated with fewer input molecules in each subsystem. 
On the other hand, we have seen that the
introduction of diffusion or exchange of molecules between partitions
enhances the variance
and reduces the range of the response, thereby reducing signaling performance.
This result is due to the presence of configurations in which the two species
are isolated from one another, compromising or even arresting signal
transmission in certain partitions. The interplay between these two effects
leads to a partition size that optimizes information transmission,
corresponding to a few molecules per partition on average,
in quantitative agreement with experiments. These effects are
generic, and hence the emergence of an optimal partition size is robust
to the specific parameters of the model.
Notably, the underlying mechanism revealed here, namely the removal of
correlations, differs fundamentally from that based on cooperativity in protein
activation, which has been argued to underlie optimal cluster size in sensory
systems \cite{Aquino11, Skoge11}.

Reversible modification reactions are ubiquitous in cell signaling, and
interactions with the cytoskeleton and lipids provide general mechanisms
for the
formation of subdomains. We therefore expect the results revealed by our study
to be applicable to a wide class of signaling systems at the membrane. We have
focused in this paper on coupled single-site modification reactions because
this
motif governs pathways specifically known to be affected by the formation of
membrane sub-domains. However, the effects we uncover also pertain to
multi-site
modification reactions, which are very common in cell signaling
\cite{Takahashi10, vanZon2007, Miller2005, Nash2001}.  Moreover, we have
focused
on systems where the reactant species are confined by a boundary which limits
diffusion.
However, similar effects could be observed in systems where proteins
are localized to raft domains, or even scaffolds or large macromolecular
complexes.
In the latter case, each
complex would effectively provide an independent reaction ``compartment,'' and
the exchange between compartments would be the result of rare dissociation
events, after which proteins could diffuse rapidly through the cytoplasm to a
different complex.
Even if the signal within each complex was not mediated via diffusive
encounters, but rather via cooperative or allosteric interactions, the
fundamental mechanism that we reveal here -- that partitioning into subsystems
removes correlations between subsystems -- remains at play.
The presence of scaffolds and macromolecular complexes
at early stages of signaling pathways is extremely common \cite{Zeke09},
suggesting that the effects discussed here are of wide biological
relevance.

\section{Methods}
\label{sec:Methods}
The CME (\ref{eq:cme}) is
solved using the method of spectral expansion \cite{Walczak09,Mugler09}.
Details of this method, the computation of mutual information,
and the spatial simulations are described in \aref{methods}.
Source code, written in MATLAB, C++, and Mathematica, used to generate all
results and figures in the main text and the {\em SI Appendix} is freely
available at {\tt http://partitioning.sourceforge.net}.

\section{Acknowledgments}
	This work is part of the research program of the ``Stichting voor
	Fundamenteel Onderzoek der Materie (FOM)'', which is financially
	supported by the ``Nederlandse organisatie voor Wetenschappelijk
	Onderzoek (NWO)''.  We thank Philippe Nghe for a critical
	reading of the manuscript.

\appendix

\section{Detailed methods}
\alabel{methods}

\subsection{Spectral solution of the master equation}
\alabel{spec}
The chemical master equation (CME) is
solved using the method of spectral expansion \cite{Walczak09,Mugler09},
described in detail in \aref{spec_detail}.  Briefly, the structure of
of the CME, in which the dynamics can be separated into two operators that
act only on $m$ or $n$ but not both, allows for its solution
to be written in the form
$p_{mn}(t)=\sum_{j=0}^M\sum_{k=0}^N
G_{jk}(t;\bar{\beta})\phi_m^j(\alpha)\phi_n^k(\bar{\beta})$,
where $\phi_m^j(\alpha)$ is the $j^{\rm th}$ eigenvector of the operator
$\mol{L}_m(\alpha)$ and similarly for $\phi_n^k(\bar{\beta})$, and
$\bar{\beta}$ is an expansion parameter on which $p_{mn}$ does not ultimately
depend. The expansion coefficients $G_{jk}(t;\bar{\beta})$ can be calculated
straightforwardly, as shown in \aref{spec_detail}. Importantly,
this spectral expansion dramatically decreases the computational complexity of
calculating $p_{mn}$: rather than solving the $(M+1)(N+1)\times(M+1)(N+1)$
system of the original CME, it is only necessary to solve $N$ linear systems
of size $(M+1)\times(M+1)$ for the vectors of coefficients $\vec{G}_k$. We
emphasize that since the system has a finite state-space, no approximations are
made in using the spectral expansion, and
the solution remains exact.
Furthermore, the moments of the steady-state
distribution $p_{mn}$ can be conveniently expressed in terms of the expansion
coefficients $G_{jk}$; in particular, $\avg{n}=G_{01}$ and
$\avg{n^2}=2G_{02}+G_{01}$. 

\subsection{Mutual information}
\alabel{mi}
The mutual information between network input and response is given by the
standard expression \cite{Shannon48}
$\I=\langle\log\{p(\alpha,n)/[p(\alpha)p(n)]\}\rangle$,
where the average is taken over the joint distribution
$p(\alpha,n)=p(n|\alpha)p(\alpha)$, and $p(n|\alpha) = \sum_{m=0}^M
p(m,n|\alpha)$ is given by the steady state of the CME. The calculation of the
mutual information requires specification of the distribution of input signals
$p(\alpha)$. We
choose $N_\alpha$ values of $\alpha$ such that $q=\alpha/(\alpha+1)=\avg{m}/M$
is uniformly-spaced over the range $0\leq q\leq 1$;
then $p(n) = \sum_{i=1}^{N_\alpha}p(n|\alpha_i)p(\alpha_i)$ and $p(\alpha_i) =
N_\alpha^{-1}$.
However, our conclusions are unaffected if we instead take a input
distribution that is unimodal or bimodal (Fig.\ \ref{fig:input}).
We take $N_\alpha>30$, for which $\I$ converges to within
$1\%$ of its large-$N_\alpha$ limit (Fig.\ \ref{fig:Nalpha}).

\subsection{Spatial simulations}
\alabel{sim}
The diffusion and reactions of $M$ $\mol{X}$ molecules and $N$ $\mol{Y}$
molecules are simulated on a two-dimensional square lattice of side length
$\lambda$ using a fixed-time-step integration scheme.  During each step of
duration $\delta t$, each particle is moved to a random neighboring lattice
site
with probability $p_D = (D/\ell^2)\delta t$, where $D$ is the diffusion
constant, and $\ell$ is the lattice spacing. Molecules have steric interactions
on the lattice, such that only one molecule can be present at each lattice site
at any time. Attempted moves to an occupied site are rejected, with the
particle
remaining at its original position. If a molecule in the $X^*$ state is
adjacent
to a molecule in the $Y$ state, the latter is converted to the $Y^*$ state with
probability $p_r = \gamma(\beta\lambda^2/M)\delta t$. To make $\pi$ partitions,
linear diffusion barriers are placed at $i\lambda/\sqrt{\pi}$ in each
direction,
where $i \in \{0, 1, \dots, \sqrt{\pi}-1 \}$.  A diffusion step which crosses
such a barrier is accepted with probability reduced by a factor $p_{\rm hop}$.
The time step $\delta t$ is chosen sufficiently small that no probability
exceeds one.

\section{Solution of the master equation by spectral expansion}
\alabel{spec_detail}

This section describes the solution via the method of spectral expansion, or the `spectral method', of the CME introduced in the main text.  The spectral method has been used fruitfully in the context of gene regulation to solve CMEs describing cascades \cite{Walczak09}, bursts \cite{Mugler09}, and oscillations \cite{Mugler10}, and a pedagogical treatment is available in \cite{Walczak12}.  Here we apply the spectral method to coupled reversible switching.

From Eqns.\ 1-2 of the main text, the stochastic dynamics of the system under study are given by the CME
\begin{equation} \label{eq:cme_supp}
	\dot{p}_{mn}=-\left[\mol{L}_m(\alpha,M)+\gamma\mol{L}_n(\beta_m,N)\right] p_{mn}, 
\end{equation}
where both operators $\mol{L}_{m}$ and $\mol{L}_n$ have the  form 
\begin{equation} \label{eq:L}
	\mol{L}_m(\alpha,M)=\alpha\left[1-{\mathbb E}_m^{-1}\right](M-m)
		+\left[1-{\mathbb E}_m^{+1}\right]m, 
\end{equation}
with ${\mathbb E}_m^{i}f(m)=f(m+i)$ defining the step operator.  The CME describes the evolution of the probability of having $m$ $\cal{X}$ proteins in the active state and $n$ $\cal{Y}$ proteins in the active state, with $\beta_m$ the coupling function by which $\cal{X}$ drives the activation of $\cal{Y}$.

\subsection{The moments do not close}

We first demonstrate that direct computation of the moments from the CME is not possible because the moments do not close.  The reason is that a nonlinearity is present in the first term of \eref{L} in the form of the factor $\beta_mn$.  As a result, the first moment depends on a higher moment, which in turn depends on an even higher moment, and so on.

To see explicitly that the moments do not close, we consider computing the dynamics of the first moment of the driven species, the mean $\avg{n}$, by summing the CME over $m$ and $n$ against $n$.  We obtain
\beq
\frac{1}{\gamma}\partial_t\avg{n} = -\avg{n} + N\avg{\beta_m} - \avg{\beta_m n},
\eeq
where averages are taken over $p_{mn}$.  We see that indeed the final term carries the nonlinearity.  Even for the simplest coupling function, i.e.\ linear coupling $\beta_m = c m$, one finds a hierarchy of moment dependencies that does not close:
\beqn
\partial_t\avg{n} &=& -\gamma\avg{n} + \gamma c N \avg{m}
	- \gamma c \avg{m n}, \\
\partial_t\avg{mn} &=& \alpha M\avg{n} - (\alpha+\gamma+1)\avg{mn}
	+ \gamma c N \avg{m^2} - \gamma c \avg{m^2n}, \\
\partial_t\avg{m^2n} &=& \dots
\eeqn
That is, the dynamics of $\avg{n}$ depend on $\avg{mn}$, whose dynamics depend on $\avg{m^2n}$, and so on.

The fact that the moments cannot be computed---indeed, not even the mean output $\avg{n}$---makes it particularly important to actually solve the CME in order to learn about the statistical properties of this system.

\subsection{The spectrum of the switch operator}

The CME is a linear equation.  Even when the rates are nonlinear functions of the molecule numbers, the CME is still linear in its degree of freedom, the joint probability.  The most straightforward way to solve a linear equation is to write its solution as an expansion in the eigenfunctions of the linear operator.  Although it is difficult to derive the eigenfunctions of the coupled operator $\L_m(\alpha,M)+\gamma\L_n(\beta_m,N)$, it is straightforward to derive the eigenfunctions of the uncoupled operator $\L_m(\alpha,M)$, which we call the switch operator.  Indeed, we will see that expanding the joint probability in eigenfunctions of the uncoupled operator greatly simplifies the form of the CME, yielding an exact solution in terms of matrix algebra.

The switch operator governs the CME for the first species ${\cal X}$; explicitly,
\beq
\elabel{Lx2}
\dot{p}_m = -\L p_m = \alpha[M-(m-1)]p_{m-1} + (m+1)p_{m+1} - [\alpha(M-m)+m]p_m,
\eeq
where for notational simplicity we have taken $\L_m(\alpha,M)\rightarrow\L$.
Its eigenvalue relation is written
\beq
\L\phi_m^j = \lambda_j\phi_m^j,
\eeq
for eigenvalues $\lambda_j$ and eigenvectors $\phi_m^j$.

\subsubsection{Eigenvalues}

The matrix form of the operator $\L$ can be read directly from \eref{Lx2}:
\beq
\elabel{Lmat}
{\bf L} = 
	\begin{pmatrix}
	M\alpha & -1 &&&&& \\
	-M\alpha & (M-1)\alpha+1 & -2 &&&& \\
	& -(M-1)\alpha & (M-2)\alpha+2 & -3 &&& \\
	&& \ddots & \ddots & \ddots && \\
	&&& -3\alpha & 2\alpha+(M-2) & -(M-1) & \\
	&&&& -2\alpha & \alpha+(M-1) & -M \\
	&&&&& -\alpha & M \\
	\end{pmatrix}.
\eeq
The tridiagonal structure follows from the fact that molecule numbers only increase or decrease by one at a time.  Practically speaking, the eigenvalues can be obtained using the fact that the determinant of a tridiagonal matrix can be computed recursively.  Performing the computation for $M = 0, 1, 2, \dots$ reveals the pattern
\beq
\elabel{eval}
\lambda_j = (\alpha+1)j, \qquad j \in \{0,1,2,\dots,M\}.
\eeq
However, \eref{eval} can be derived more rigorously by making use of a generating function.  We present this derivation next, since the generating function formalism will also prove quite useful in deriving the eigenvectors and solving the CME.

The generating function is an expansion in any complete basis for which the probability distribution provides the expansion coefficients \cite{vanKampen92}.  Choosing as our basis the set of polynomials in some continuous variable $x$, the generating function is defined
\beq
\elabel{gf}
G(x) = \sum_{m=0}^M p_m x^m.
\eeq
The probability distribution is recovered via the inverse transform
\beq
\elabel{inv}
p_m = \frac{1}{m!}\partial_x^m[G(x)]_{x=0}.
\eeq
A key utility of the generating function is turning the CME, which is a set of ordinary differential equations (ODEs), into a single partial differential equation.  Indeed, summing \eref{Lx2} against $x^m$ yields
\beq
\elabel{G}
\dot{G} = -(x-1)[(\alpha x+1)\partial_x - \alpha M] G,
\eeq
where the appearances of $x$ and $\partial_x$ arise from the shifts $m-1$ and $m+1$, respectively.
\eref{G} directly gives the form of the operator in $x$ space: $\L = (x-1)[(\alpha x+1)\partial_x - \alpha M]$.  The eigenfunctions are then obtained from the relation $\L\phi^j(x) = \lambda_j \phi^j(x)$ by separating variables and integrating:
\beq
\elabel{phix}
\phi^j(x) = (\alpha+1)^{-M}(x-1)^{\lambda_j/(\alpha+1)}(\alpha x+1)^{M-\lambda_j/(\alpha+1)}.
\eeq
Here, the constant factor $(\alpha+1)^{-M}$ is determined by application of the normalization condition $G(1) = 1$ to the steady state solution, which is obtained by setting $\lambda_j=0$:
\beq
\elabel{ss}
G(x) = \left( \frac{\alpha x+1}{\alpha+1} \right)^M.
\eeq
We will solve \eref{G} in two ways: by the method of characteristics and by expansion in the eigenfunctions; together these solutions will reveal the eigenvalues.

First, the method of characteristics \cite{Farlow93} posits that the dependence of $G$ on $x$ and $t$ occurs via some parametric variable $s$, i.e.\ $G(x,t) = G[x(s),t(s)]$.  The chain rule then gives $dG/ds = (\partial G/\partial x)(dx/ds) + (\partial G/\partial t)(dt/ds)$, which when compared term by term with \eref{G} yields three ordinary differential equations:
\beq
\frac{dt}{ds} = 1, \qquad \frac{dx}{ds} = (x-1)(\alpha x+1), \qquad \frac{dG}{ds} = \alpha M(x-1).
\eeq
The first identifies $s=t$, with which the second is solved by
\beq
\elabel{char}
z = \frac{x-1}{\alpha x+1}e^{-(\alpha +1)t},
\eeq
where $z$ is a constant of integration.  The crux of the method is that \eref{char} defines a characteristic curve on which $G$ must depend, i.e.\ $G(x,t) = f[z(x,t)]g(x,t)$, where $f$ and $g$ are unknown functions, and $z$ has been promoted to a characteristic function of $x$ and $t$.  The function $g$ is identified by realizing that steady state is reached as $t\rightarrow\infty$, for which $f(z) \rightarrow f(0)$ no longer depends on $x$ or $t$.  Therefore, $g$ must be the steady state function given in \eref{ss}:
\beq
G(x,t) = \left( \frac{\alpha x+1}{\alpha+1} \right)^M f(z).
\eeq
Although we still do not know $f$, we may Taylor expand it around the point $z=0$, yielding
\beq
\elabel{Gform1}
G(x,t) = \left( \frac{\alpha x+1}{\alpha+1} \right)^M \sum_{j=0}^\infty c_j z^j
	= \left( \frac{\alpha x+1}{\alpha+1} \right)^M
		\sum_{j=0}^\infty c_j \left( \frac{x-1}{\alpha x+1} \right)^j e^{-(\alpha +1)jt},
\eeq
where $c_j \equiv \partial_z^j[f(z)]_{z=0}/j!$.

Second, because \eref{G} is linear, we may also write down its solution as an expansion in the eigenfunctions of its linear operator:
\beq
\elabel{Geig}
G(x,t) = \sum_j C_j(t) \phi^j(x).
\eeq
Under the assumption that the eigenfunctions are orthogonal (which will be shown in the next section), inserting \eref{Geig} into \eref{G} yields an independent ODE for each $C_j$, $\dot{C}_j = -\lambda_jC_j$, which is solved by $C_j(t) = c_j e^{-\lambda_j t}$ for initial conditions $c_j$.  Inserting this functional form and that for $\phi_j(t)$ (\eref{phix}) into \eref{Geig} yields
\beq
\elabel{Gform2}
G(x,t) = \left( \frac{\alpha x+1}{\alpha+1} \right)^M
		\sum_j c_j \left( \frac{x-1}{\alpha x+1} \right)^{\lambda_j/(\alpha+1)} e^{-\lambda_jt}.
\eeq
Comparison of Eqns.\ \ref{eq:Gform1} and \ref{eq:Gform2} reveals both the expression for the eigenvalues, $\lambda_j = (\alpha+1)j$, and a limit on their domain, the nonnegative integers $j \in \{ 0, 1, 2, \dots, \infty \}$.  Of course, the domain can be a subset of the nonnegative integers; then some $c_j$ in \eref{Gform2} would be zero.  Indeed, since ${\bf L}$ is a finite matrix of size $M+1$ by $M+1$ (\eref{Lmat}), it is spanned by $M+1$ linearly independent eigenvectors, meaning we expect only $M+1$ eigenvalues.  In fact, the only set of $M+1$ nonnegative integers that satisfies the requirement that the trace of ${\bf L}$, $\sum_{m=0}^M [(M-m)\alpha + m] = (\alpha+1)M(M+1)/2$, equals the sum of the eigenvalues, $\sum_j (\alpha+1)j$, is $j \in \{0, 1, 2, \dots, M\}$.  Thus, we arrive at the result
\beq
\elabel{eval2}
\lambda_j = (\alpha+1)j, \qquad j \in \{0,1,2,\dots,M\},
\eeq
as proposed by inspection in \eref{eval}.

\subsubsection{State space notation}

The linear algebraic manipulations we have done thus far can be cast in the more abstract notation of state spaces, commonly used in quantum mechanics \cite{Townsend00}.  We will find this notation useful in later sections, for example in transforming between the molecule number basis and the eigenbasis.  Specifically, we introduce a state $\ket{p}$ that can be projected into $\bra{m}$ space to give the probability distribution, or into $\bra{x}$ space to give the generating function:
\beq
\ip{m}{p} = p_m, \qquad \ip{x}{p} = G(x).
\eeq
In the same way, the $j$th eigenstate $\ket{j}$ is projected into $\bra{m}$ space to give the $j$th eigenvector, or into $\bra{x}$ space to give the $j$th eigenfunction:
\beq
\ip{m}{j} = \phi_m^j, \qquad \ip{x}{j} = \phi^j(x).
\eeq
This notation offers new insight into our definition of the generating function.
For example, Eqn.\ \ref{eq:gf} can now be written
\beq
\elabel{insertm}
\ip{x}{p} = \sum_{m=0}^M \ip{x}{m}\ip{m}{p},
\eeq
where we have recognized
\beq
\ip{x}{m} = x^m
\eeq
as the projection of the state $\ket{m}$ into $\bra{x}$ space.  Eqn.\ \ref{eq:insertm} has a clear interpretation: we have inserted a complete set of $\ket{m}$ states.  Similarly, Eqn.\ \ref{eq:inv} can now be written
\beq
\elabel{insertx}
\ip{m}{p} = \oint \d x \frac{G(x)}{x^{m+1}} = \oint \d x \ip{m}{x}\ip{x}{p}.
\eeq
In the first step, we have rewritten Eqn.\ \ref{eq:inv} using Cauchy's theorem, where $\d x \equiv dx/2\pi i$, and the contour surrounds the pole at $x=0$.  In the second step, we have recognized
\beq
\ip{m}{x} = \frac{1}{x^{m+1}}
\eeq
as the conjugate to $\ip{x}{m}$.  Eqn.\ \ref{eq:insertx} has the clear interpretation of inserting a complete set of $\ket{x}$ states, under an inner product defined by the complex integration.  The choice of inner product and of conjugate state are made such that orthonormality is preserved, a fact which we may confirm by again employing Cauchy's theorem:
\beq
\elabel{ortho}
\ip{m}{m'} = \oint \d x \ip{m}{x}\ip{x}{m'} = \oint \d x \frac{x^{m'}}{x^{m+1}}
	= \frac{1}{m!} \partial_x^m \left[ x^{m'} \right]_{x=0} \theta(m>0) = \delta_{mm'}.
\eeq
Finally, the dynamics in \eref{G} can be written in state space as
\beq
\elabel{pdot}
\ket{\dot{p}} = -\Lh \ket{p} = -(\ap -1) [(\alpha\ap +1)\am-\alpha M]\ket{p},
\eeq
where we have defined the operators $\ap$ and $\am$ whose projections in $x$ space are $\bra{x}\ap = x$ and $\bra{x}\am = \partial_x$.  These are analogous to the raising and lowering operators in the well known treatment of the quantum harmonic oscillator.  This operator formalism for the generating function was first developed in the 1970s; for a review see \cite{Mattis98}.

\subsubsection{Eigenvectors}

The state space notation facilitates a derivation of the functional form of the eigenvectors:
\beq
\elabel{evec0}
\phi_m^j = \ip{m}{j} = \oint \d x \ip{m}{x} \ip{x}{j}
	= \oint \d x \frac{1}{x^{m+1}}\frac{(x-1)^j(\alpha x+1)^{M-j}}{(\alpha+1)^M}.
\eeq
Here we have inserted the eigenfunctions
\beq
\elabel{jx}
\phi_j(x) = \ip{x}{j} = \frac{(x-1)^j(\alpha x+1)^{M-j}}{(\alpha+1)^M}
\eeq
from \eref{phix}, with eigenvalues given by \eref{eval2}.  We use Cauchy's theorem to perform the integration and recognize that derivatives of a product follow a binomial expansion:
\beqn
\phi_m^j &=& \frac{1}{(\alpha+1)^M} \frac{1}{m!} \partial_x^m \left[(\alpha x+1)^{M-j}(x-1)^j\right]_{x=0} \\
&=& \frac{1}{(\alpha+1)^M} \frac{1}{m!} \sum_{\ell=0}^m {m \choose \ell}
	\partial_x^\ell \left[ (\alpha x+1)^{M-j} \right]_{x=0} \partial_x^{m-\ell} \left[ (x-1)^j \right]_{x=0} \\
&=& \frac{1}{(\alpha+1)^M} \frac{1}{m!} \sum_{\ell=0}^m \frac{m!}{(m-\ell)!\ell!}
	\left[ \frac{(M-j)! \alpha^\ell}{(M-j-\ell)!} \theta(\ell \le M-j) \right]
	\left[ \frac{j! (-1)^{j-m+\ell}}{(j-m+\ell)!} \theta(m-\ell \le j) \right] \\
\elabel{evec}
&=& \frac{(-1)^{j-m}}{(\alpha+1)^M} \sum_{\ell \in \Omega} {M-j \choose \ell} {j \choose m-\ell} (-\alpha)^\ell.
\eeqn
Here the domain $\Omega$ results from the derivatives and is defined by $\max(0,m-j) \le \ell \le \min(m,M-j)$.  \eref{evec} gives the expression for the eigenvectors.  For $j=0$ the expression reduces to the binomial distribution in terms of the occupancy $q = \alpha/(\alpha+1)$, as it must, since this is the steady state of the uncoupled process:
\beq
\elabel{binomial}
\phi_m^0 = {M \choose m} \frac{\alpha^m}{(\alpha+1)^M} = {M \choose m} q^m (1-q)^{M-m}.
\eeq
This function has one maximum, and in general the $j$th eigenvector has $j+1$ extrema, making the eigenvectors qualitatively similar to Fourier modes or eigenfunctions of the quantum harmonic oscillator.

The switch operator $\Lh$ is not Hermitian.  A consequence is that its conjugate eigenvectors $\psi_m^j = \ip{j}{m}$ (row vectors) are not complex conjugates of its eigenvectors $\phi_m^j = \ip{m}{j}$ (column vectors).  Rather, they are distinct functions that must be constructed to obey an orthonormality relation in order to constitute a complete basis.  The orthonormality relation can be used to derive their form in $x$ space, $\psi^j(x) = \ip{j}{x}$:
\beq
\elabel{ortho2}
\delta_{jj'} = \ip{j}{j'} = \oint \d x \ip{j}{x} \ip{x}{j'}
	= \oint \d x\ \psi^j(x) \frac{(x-1)^{j'}(\alpha x+1)^{M-{j'}}}{(\alpha+1)^M}
	= \oint \d z_0\ z_0^{j'} f_j(z_0).
\eeq
Here we have defined $z_0 \equiv (x-1)/(\alpha x+1)$ and $f_j(z_0) \equiv \psi^j(x)(\alpha x+1)^{M+2}/(\alpha+1)^{M+1}$ in order to draw an equivalence between \eref{ortho2} and \eref{ortho}, which then implies $f_j(z_0) = 1/z_0^{j+1} = (\alpha x+1)^{j+1}/(x-1)^{j+1}$, or
\beq
\elabel{psij}
\psi^j(x) = \frac{(\alpha+1)^{M+1}}{(\alpha x+1)^{M-j+1}(x-1)^{j+1}}.
\eeq
\eref{psij} gives the form of the conjugate eigenfunctions in $x$ space, which can be used to derive the expression for the conjugate eigenvectors as in \erefn{evec0}{evec}:
\beqn
\psi_m^j &=& \ip{j}{m} = \oint \d x \ip{j}{x} \ip{x}{m}
	= \oint \d x \frac{(\alpha+1)^{M+1}}{(\alpha x+1)^{M-j+1}(x-1)^{j+1}} x^m \\
\elabel{evec2}
	&=& \sum_{\ell \in \Omega} {M-j+\ell \choose \ell} {m \choose j-\ell} (-\alpha)^\ell (\alpha+1)^{j-\ell}.
\eeqn
Here $\Omega$ is defined by $\max(0,j-m)\le\ell\le j$.  \eref{evec2} gives the expression for the conjugate eigenvectors.  They are $j$th order polynomials in $m$.

\subsection{Expanding the coupled problem in uncoupled eigenfunctions}

We now solve the CME by expanding the solution in the eigenfunctions of the uncoupled operator.  This procedure is most easily done in state space, in which the CME reads
\beq
\ket{\dot{p}} = -[\Lh_x(\alpha) + \gamma\Lh_{xy}]\ket{p}
\eeq
where
\beqn
\Lh_x(\alpha) &=& (\ap_x -1) [(\alpha\ap_x +1)\am_x-\alpha M], \\
\Lh_{xy} &=& (\ap_y -1) [(\bh_x\ap_y +1)\am_y-\bh_x N],
\eeqn
as in \eref{pdot}, and we have introduced the operator $\bh_x$ whose action on the state $\ket{m}$ yields the coupling function, $\bh_x\ket{m} = \beta_m\ket{m}$.  The first step is to write the full operator as two uncoupled operators plus a correction term.  Introducing the constant $\bb$ to parameterize the second uncoupled operator, the CME becomes
\beq
\ket{\dot{p}} = -[\Lh_x(\alpha) + \gamma\Lh_y(\bb) + \gamma\Gh_x\Dh_y]\ket{p}
\eeq
where we have explicitly denoted the fact that the correction term $\Lh_{xy} - \Lh_y(\bb)$ factorizes into two operators that act on each of the $x$ and $y$ sectors alone:
\beqn
\Gh_x &\equiv& \bh_x-\bb, \\
\Dh_y &\equiv& (\ap_y-1)(\ap_y\am_y-N).
\eeqn
The second step is to expand the solution in the eigenfunctions of the two uncoupled operators.  Introducing $k$ as the mode index for the eigenstates of $\Lh_y(\bb)$, we write
\beq
\elabel{expand}
\ket{p} = \sum_{j=0}^M\sum_{k=0}^N G_{jk} \ket{j,k}.
\eeq
Inserting this form into the CME, projecting with the conjugate state $\bra{j',k'}$, and summing over $j$ and $k$ yields the dynamics for the expansion coefficients $G_{jk}$:
\beq
\dot{G}_{jk} = -[(\alpha+1)j+\gamma(\bb+1)k]G_{jk}
	-\gamma\sum_{j'=0}^M \Gamma_{jj'} \sum_{k'=0}^N \Delta_{kk'} G_{j'k'}.
\eeq
Here the first term is diagonal and reflects the actions of the uncoupled operators on their eigenstates.  The second term contains the corrections $\Gamma_{jj'} = \bra{j}\Gh_x\ket{j'}$ and $\Delta_{kk'} = \bra{k}\Dh_y\ket{k'}$.  The first correction is directly evaluated by inserting a complete set of $m$ states:
\beqn
\Gamma_{jj'} &=& \sum_{m=0}^M \bra{j}(\bh_x-\bb)\ket{m}\ip{m}{j'}
	= \sum_{m=0}^M \ip{j}{m}(\beta_m-\bb)\ip{m}{j'} \\
\elabel{Gamma}
	&=& \sum_{m=0}^M \psi_m^j(\beta_m-\bb)\phi_m^{j'}.
\eeqn
We see that $\Gamma_{jj'}$ is the simply the difference between the coupling function and the constant parameter, rotated into eigenspace.  Notably, for linear coupling, $\Gamma_{jj'}$ is tridiagonal (see \sref{aux}).  The second correction is most easily evaluated by inserting a complete set of $y$ states; the result, derived in \sref{aux}, is
\beq
\elabel{Delta}
\Delta_{kk'} = k \delta_{kk'} - (N-k+1) \delta_{k-1,k'}.
\eeq
We see that $\Delta_{kk'}$ is subdiagonal in $k$, which simplifies the dynamics of $G_{jk}$ to
\beq
\elabel{Gmat}
\dot{G}_{jk} = -\sum_{j=0}^M \Lambda^k_{jj'} G_{j'k} + \gamma (N-k+1) \sum_{j=0}^M \Gamma_{jj'} G_{j',k-1},
\eeq
where we define the matrix acting on the diagonal part as
\beq
\elabel{Lambda}
\Lambda^k_{jj'} \equiv [(\alpha+1)j+\gamma(\bb+1)k]\delta_{jj'}+\gamma k \Gamma_{jj'}.
\eeq
The subdiagonality allows one to write the steady state of \eref{Gmat} as an iterative scheme, by which the $k$th column of $G_{jk}$ is computed from the $(k-1)$th column:
\beq
\elabel{iter}
\vec{G}_k = \gamma(N-k+1){\bf \Lambda}_k^{-1} {\bf \Gamma} \vec{G}_{k-1}.
\eeq
The scheme is initialized with
\beq
\elabel{IC}
\vec{G}_0 = \delta_{j0}
\eeq
(see \sref{aux}), and the joint distribution is recovered via
\beq
\elabel{psol}
p_{mn} = \sum_{j=0}^M \sum_{k=0}^N G_{jk} \phi_m^j \phi_n^k,
\eeq
which is the projection of \eref{expand} into $\bra{m,n}$ space.

\eref{psol} constitutes an exact steady state solution to the CME, with $G_{jk}$ computed iteratively via \erefs{iter}{IC}, auxiliary matrices defined in \erefs{Gamma}{Lambda}, and the eigenvectors given by \erefs{evec}{evec2}.  Importantly, the computational complexity of the solution has been dramatically reduced: rather than solving the original CME (\eref{cme_supp}), which requires inverting its operator of size $(M+1)(N+1) \times (M+1)(N+1)$, \eref{iter} makes clear that it is only necessary to invert $N$ smaller matrices of size $(M+1)\times(M+1)$, i.e.\ the matrices ${\bf \Lambda}_k$ for $k\in\{1,2,\dots,N\}$.

\subsection{Exact expressions for moments}

Now that we have an exact solution to the CME in terms of a spectral expansion, moments take  an exact form in terms of the expansion coefficients.  We thus circumvent the problem of moment closure, instead arriving at compact expressions that require only the inversion and multiplication of finite matrices via \eref{iter}.

Moments are most easily computed from the generating function, $G(x,y)$.  For example, the $\nu$th moment of the output is
\beq
\avg{n^\nu} = \left[ \left( y \partial_y \right)^\nu G(x=1,y) \right]_{y=1}.
\eeq
In terms of the expansion, the generating function is $G(x,y) = \ip{x,y}{p} = \sum_{j=0}^M\sum_{k=0}^N G_{jk} \ip{x}{j} \ip{y}{k}$, and using the fact that $\ip{x=1}{j} = \delta_{j0}$ (\eref{jx}), we have
\beq
\avg{n^\nu} = \sum_{k=0}^N G_{0k} \left[ \left( y \partial_y \right)^\nu \ip{y}{k} \right]_{y=1}.
\eeq
Inserting the expression for $\ip{y}{k}$ (\eref{jx}) and defining $w \equiv \log y$, we obtain
\beq
\elabel{mombb}
\avg{n^\nu} = \sum_{k=0}^N G_{0k}
	\partial_w^\nu \left[ \frac{(e^w-1)^k(\bb e^w+1)^{N-k}}{(\bb+1)^N} \right]_{w=0}.
\eeq
At this point we recall that $\bb$ is a constant we introduce to parameterize the expansion.  The expression for the moments therefore cannot depend on $\bb$: if we change $\bb$, the expression in brackets changes, but the expansion coefficients $G_{0k}$ also change, such that \eref{mombb} evaluates to the same $\bb$-independent form.  We are therefore free to set $\bb$ to any value, and the choice $\bb=0$ makes the derivative easiest to evaluate.  Thus we have
\beq
\avg{n^\nu} = \sum_{k=0}^N G_{0k}
	\partial_w^\nu \left[ (e^w-1)^k \right]_{w=0},
\eeq
where it is now understood that $G_{0k}$ is computed with $\bb=0$.  Evaluating the derivative yields
\beqn
\avg{n^\nu} &=& \sum_{k=0}^N G_{0k}
	\left[ \sum_{\ell=1}^{\min(k,\nu)} \stir{\nu}{\ell} \frac{k!}{(k-\ell)!}
		e^{\ell w} (e^w-1)^{k-\ell} \right]_{w=0} \\
	&=& \sum_{k=0}^N G_{0k}
	\sum_{\ell=1}^{\min(k,\nu)} \stir{\nu}{\ell} \frac{k!}{(k-\ell)!} \delta_{k\ell} \\
	&=& \sum_{k=1}^{\min(\nu,N)} G_{0k} \stir{\nu}{k} k!
\eeqn
in terms of the Stirling numbers of the second kind,
\beq
\stir{\nu}{k} = \frac{1}{k!} \sum_{\ell=0}^k (-1)^{k-\ell} {k \choose \ell} \ell^\nu.
\eeq
For example, the first moment, second moment, and variance are
\beqn
\avg{n} &=& G_{01}, \\
\avg{n^2} &=& G_{01}+2G_{02}, \\
\elabel{noise_supp}
\sigma_n^2 &=& \avg{n^2} - \avg{n}^2 = G_{01}+2G_{02}-G_{01}^2.
\eeqn
These are exact expressions for the moments in terms of the expansion coefficients $G_{0k}$, which are obtained by matrix inversion and multiplication via \eref{iter}, e.g.\ in Mathematica.

An informative special case is immediately revealed when $N=1$, for which $G_{02}$ does not exist, i.e.\ $\avg{n} = G_{01}$ and $\sigma_n^2 = G_{01}-G_{01}^2$, or
\beq
\elabel{switch}
\sigma_n^2 = \avg{n} (1-\avg{n}) \qquad (N=1).
\eeq
Here there is only one output molecule.  The relationship between its mean activation and the associated noise must therefore obey the known result for a single binary switch, \eref{switch}.

\subsection{Auxiliary calculations}
\slabel{aux}

Here we show that $\Gamma_{jj'}$ is tridiagonal for linear $\beta_m = cm$:
\beqn
\Gamma_{jj'} &=& \bra{j} \Gh_x \ket{j'} \\
\elabel{Gamma1}
	&=& \bra{j} (c\ap_x\am_x-\bb) \ket{j'} \\
	&=& -\bb\delta_{jj'} + c\oint \d x \ip{j}{x}\bra{x} \ap_x\am_x \ket{j'} \\
	&=& -\bb\delta_{jj'} + c\oint \d x \ip{j}{x} x\partial_x \ip{x}{j'} \\
	&=& -\bb\delta_{jj'} + c\oint \d x \ip{j}{x} x\partial_x
		\frac{(x-1)^{j'}(\alpha x+1)^{M-j'}}{(\alpha+1)^M} \\
	&=& -\bb\delta_{jj'} + c\oint \d x \ip{j}{x} \frac{x}{(\alpha+1)^M}
		\left[ j'(x-1)^{j'-1}(\alpha x+1)^{M-j'} \right. \nonumber \\
	&&	\qquad \qquad \qquad \qquad \quad \quad \quad \quad \quad
		\left . + (x-1)^{j'}(M-j')(\alpha x+1)^{M-j'-1}\alpha \right] \\
	&=& -\bb\delta_{jj'} + c\oint \d x \ip{j}{x} \frac{x(x-1)^{j'-1}(\alpha x+1)^{M-j'-1}}{(\alpha+1)^M}
		\left[ j'(\alpha x+1)+(x-1)(M-j')\alpha \right] \\
\elabel{Gamma2}
	&=& -\bb\delta_{jj'} + \frac{c}{\alpha+1}\oint \d x \ip{j}{x}
		\frac{x(x-1)^{j'-1}(\alpha x+1)^{M-j'-1}}{(\alpha+1)^M}
		\left\{ j'(\alpha x+1)^2  \right. \nonumber \\
	&&	\qquad \qquad \qquad \qquad \qquad \qquad \qquad \qquad \qquad \qquad \qquad \qquad
		\left . + [\alpha(M-j')+j'](\alpha x+1)(x-1)  \right. \nonumber \\
	&&	\qquad \qquad \qquad \qquad \qquad \qquad \qquad \qquad \qquad \qquad \qquad \qquad
		\left . + \alpha(M-j')(x-1)^2 \right\} \\
	&=& -\bb\delta_{jj'} + \frac{c}{\alpha+1}\oint \d x \ip{j}{x}
		\left\{ \frac{(x-1)^{j'-1}(\alpha x+1)^{M-j'+1}}{(\alpha+1)^M} \left[ j' \right] \right. \nonumber \\
	&&	\qquad \qquad \qquad \qquad \quad \quad \quad \quad 
		\left .	+ \frac{(x-1)^{j'}(\alpha x+1)^{M-j'}}{(\alpha+1)^M}
			\left[ \alpha(M-j')+j' \right] \right. \nonumber \\
	&&	\qquad \qquad \qquad \qquad \quad \quad \quad \quad 
		\left . + \frac{(x-1)^{j'-1}(\alpha x+1)^{M-j'+1}}{(\alpha+1)^M} \left[ \alpha(M-j') \right] \right\} \\
	&=& -\bb\delta_{jj'} + \frac{c}{\alpha+1}\oint \d x \ip{j}{x}
		\left\{ \ip{x}{j'-1} j' + \ip{x}{j'} \left[ \alpha(M-j')+j' \right] + \ip{x}{j'+1} \alpha(M-j') \right\} \\
	&=& -\bb\delta_{jj'} + \frac{c}{\alpha+1}
		\left\{ \ip{j}{j'-1} j' + \ip{j}{j'} \left[ \alpha(M-j')+j' \right] + \ip{j}{j'+1} \alpha(M-j') \right\} \\
	&=& \frac{cj'}{\alpha+1}\delta_{j,j'-1}
		+ \left\{ \frac{c[\alpha(M-j')+j']}{\alpha+1} - \bb \right\} \delta_{jj'}
		+ \frac{c\alpha(M-j')}{\alpha+1} \delta_{j,j'+1}.
\eeqn
\eref{Gamma1} recognizes that $\bh_x=c\ap_x\am_x$ is the operator representation of $\beta_m$ (since $\ap\am$ is the number operator, i.e.\ $\ap_x\am_x\ket{m} = m\ket{m}$), and \eref{Gamma2} uses the algebraic fact that $x[j'(\alpha x+1)+(x-1)(M-j')\alpha](\alpha+1) = j'(\alpha x+1)^2 + [\alpha(M-j')+j'](\alpha x+1)(x-1) + \alpha(M-j')(x-1)^2$, which is straightforward to verify.

Here we derive \eref{Delta}:
\beqn
\Delta_{kk'} &=& \bra{k} \Dh_y \ket{k'} \\
	&=& \bra{k} (\ap_y-1)(\ap_y\am_y-N) \ket{k'} \\
	&=& \oint \d y \ip{k}{y}\bra{y} (\ap_y-1)(\ap_y\am_y-N) \ket{k'} \\
	&=& \oint \d y \ip{k}{y} (y-1)(y\partial_y-N) \ip{y}{k'} \\
	&=& \oint \d y \ip{k}{y} (y-1)(y\partial_y-N) \frac{(y-1)^{k'}(\bb y+1)^{N-k'}}{(\bb+1)^N} \\
	&=& \oint \d y \ip{k}{y} \frac{(y-1)}{(\bb+1)^N}
		\left[ yk'(y-1)^{k'-1}(\bb y+1)^{N-k'} + y(y-1)^{k'}(N-k')(\bb y+1)^{N-k'-1}\bb \right. \nonumber \\
	&&	\qquad \qquad \qquad \qquad \quad \left . -N(y-1)^{k'}(\bb y+1)^{N-k'} \right] \\
	&=& \oint \d y \ip{k}{y} \frac{(y-1)^{k'}(\bb y+1)^{N-k'-1}}{(\bb+1)^N}
		\left[ yk'(\bb y+1) + y(y-1)(N-k')\bb \right. \nonumber \\
	&&	\qquad \qquad \qquad \qquad \qquad \qquad \qquad \quad\ \ \left. -N(y-1)(\bb y+1) \right] \\
	&=& \oint \d y \ip{k}{y} \frac{(y-1)^{k'}(\bb y+1)^{N-k'-1}}{(\bb+1)^N}
		\left[ k'(\bb y+1) - (y-1)(N-k') \right] \\
	&=& \oint \d y \ip{k}{y} \left[ k' \frac{(y-1)^{k'}(\bb y+1)^{N-k'}}{(\bb+1)^N}
		 - (N-k') \frac{(y-1)^{k'+1}(\bb y+1)^{N-(k'+1)}}{(\bb+1)^N} \right] \\
	&=& \oint \d y \ip{k}{y} \left[ k' \ip{y}{k'} - (N-k') \ip{y}{k'+1} \right] \\
	&=& k' \ip{k}{k'} - (N-k') \ip{k}{k'+1} \\
	&=& k' \delta_{kk'} - (N-k') \delta_{k,k'+1} \\
	&=& k \delta_{kk'} - (N-k+1) \delta_{k-1,k'}.
\eeqn

Here we derive \eref{IC}:
\beqn
\vec{G}_0 &=& G_{j0} \\
	&=& \ip{j,k=0}{p} \\
	&=& \sum_{m=0}^M \sum_{n=0}^N \ip{j}{m} \ip{k=0}{n} \ip{m,n}{p} \\
\elabel{IC1}
	&=& \sum_{m=0}^M \sum_{n=0}^N \ip{j}{m} p_{mn} \\
	&=& \sum_{m=0}^M \ip{j}{m} p_m \\
\elabel{IC2}
	&=& \sum_{m=0}^M \ip{j}{m} \ip{m}{j=0} \\
	&=& \ip{j}{j=0} \\
	&=& \delta_{j0}.	
\eeqn
\eref{IC1} uses \eref{evec2} to obtain $\ip{k=0}{n} = 1$, and \eref{IC2} recognizes that $p_m$ is the steady state of the uncoupled operator, $p_m = \phi_m^0 = \ip{m}{j=0}$.

\clearpage
\section{Supplementary figures}
\alabel{fig}

\vspace{.5in}
\begin{figure}[h]
	\begin{center}
		\includegraphics[width=0.75\textwidth]{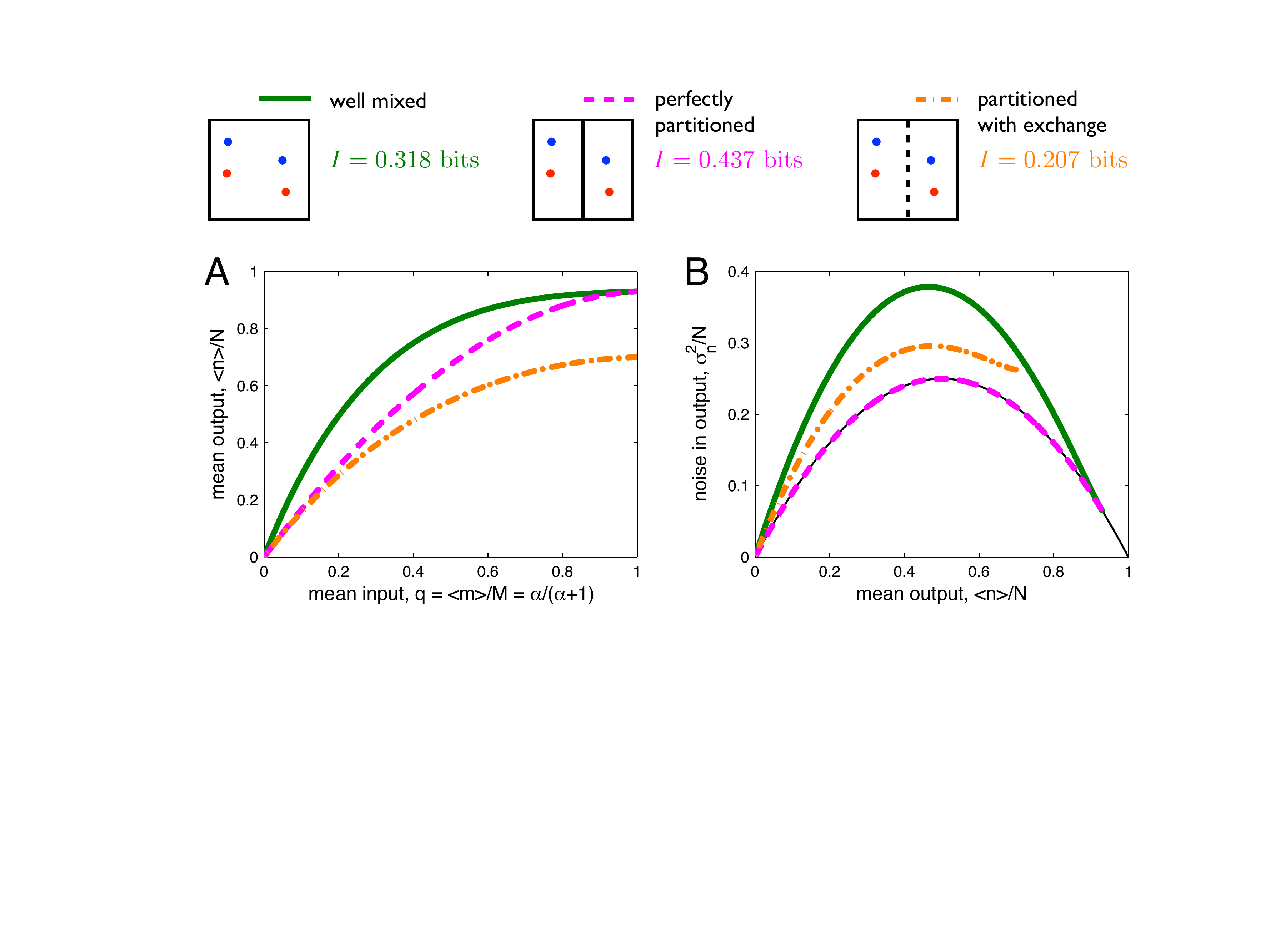}
		\line(1,0){350}\\
		\includegraphics[width=0.75\textwidth]{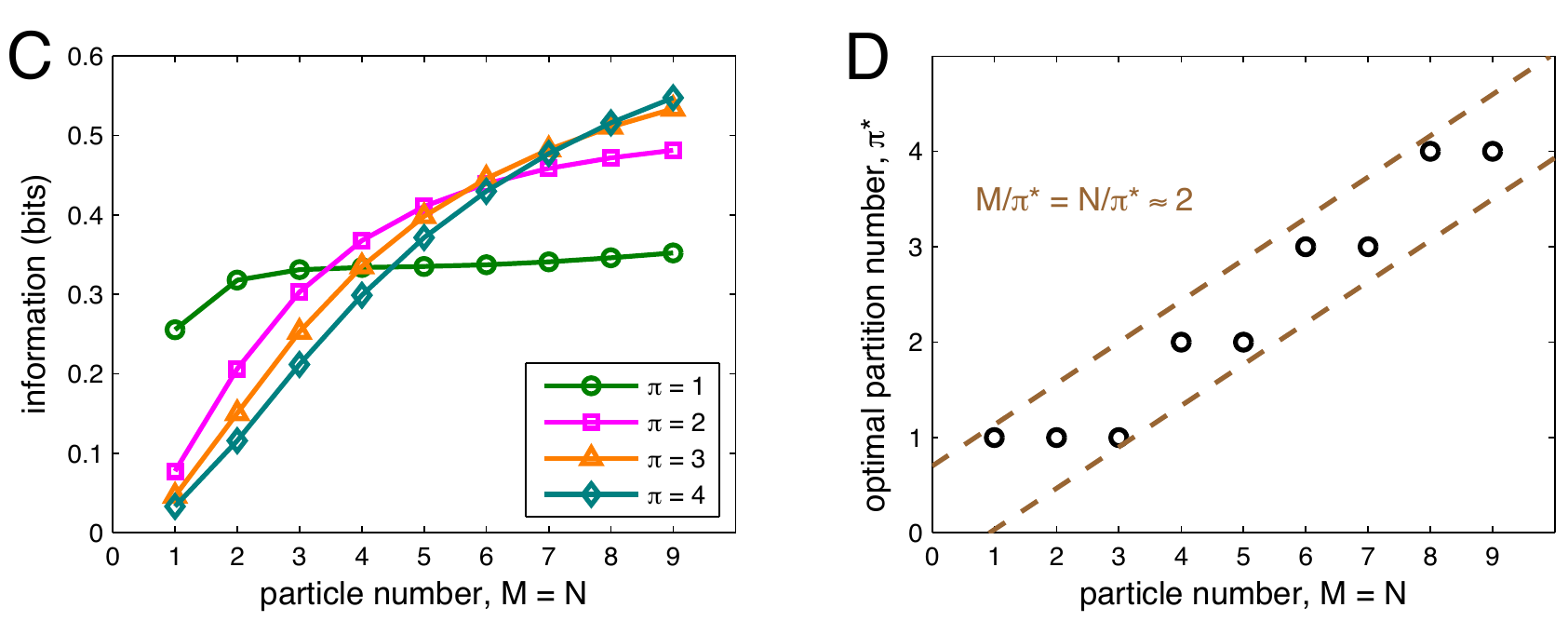}
	\end{center}
	\caption{	\label{fig:mm}
		The effects of partitioning persist for Michaelis-Menten coupling.
		The coupling is described by
		$\beta_m^{(i)} = \beta m_i/[m_i+(V/\pi)K] = \beta m_i/(m_i+\phi M/\pi)$,
		where $m_i$ is the number of $X^*$ molecules in partition $i \in \{1,\dots,\pi\}$,
		and $\phi \equiv KV/M$ is a constant.  Here $\beta=20$, $\phi = 1/2$, and $\gamma=1$.
	}
\end{figure}

		{\bf A, B} As in Fig.\ 2 of the main text, with $M=N=2$,
		perfect partitioning linearizes the input-output relation and reduces the noise,
		transmitting more information than the well-mixed system;
		further, allowing exchange among partitions
		compresses the response and
		increases the noise compared to the perfectly partitioned system,
		transmitting less information than the well-mixed system.
		
		{\bf C, D} As in Fig.\ 5 of the main text,
		an information-optimal partition size,
		here $M/\pi^* = N/\pi^* \approx 2$, emerges
		due to the trade-off between optimizing signaling reliability
		and avoiding unfavorable configurations.

\clearpage

\ \vspace{1in}
\begin{figure}[h]
	\begin{center}
		\includegraphics[width=0.45\textwidth]{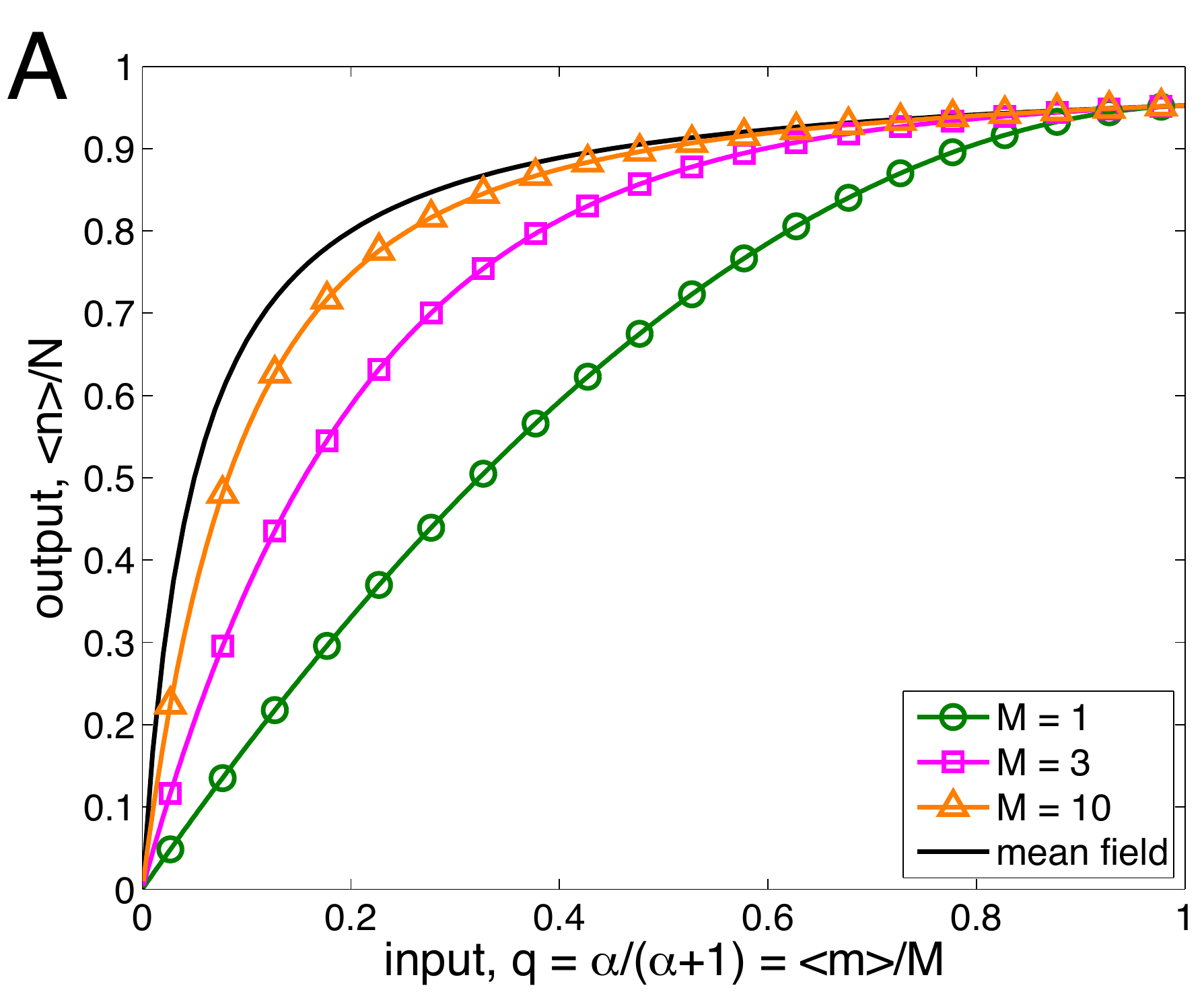}
		\hspace{.05\textwidth}
		\includegraphics[width=0.45\textwidth]{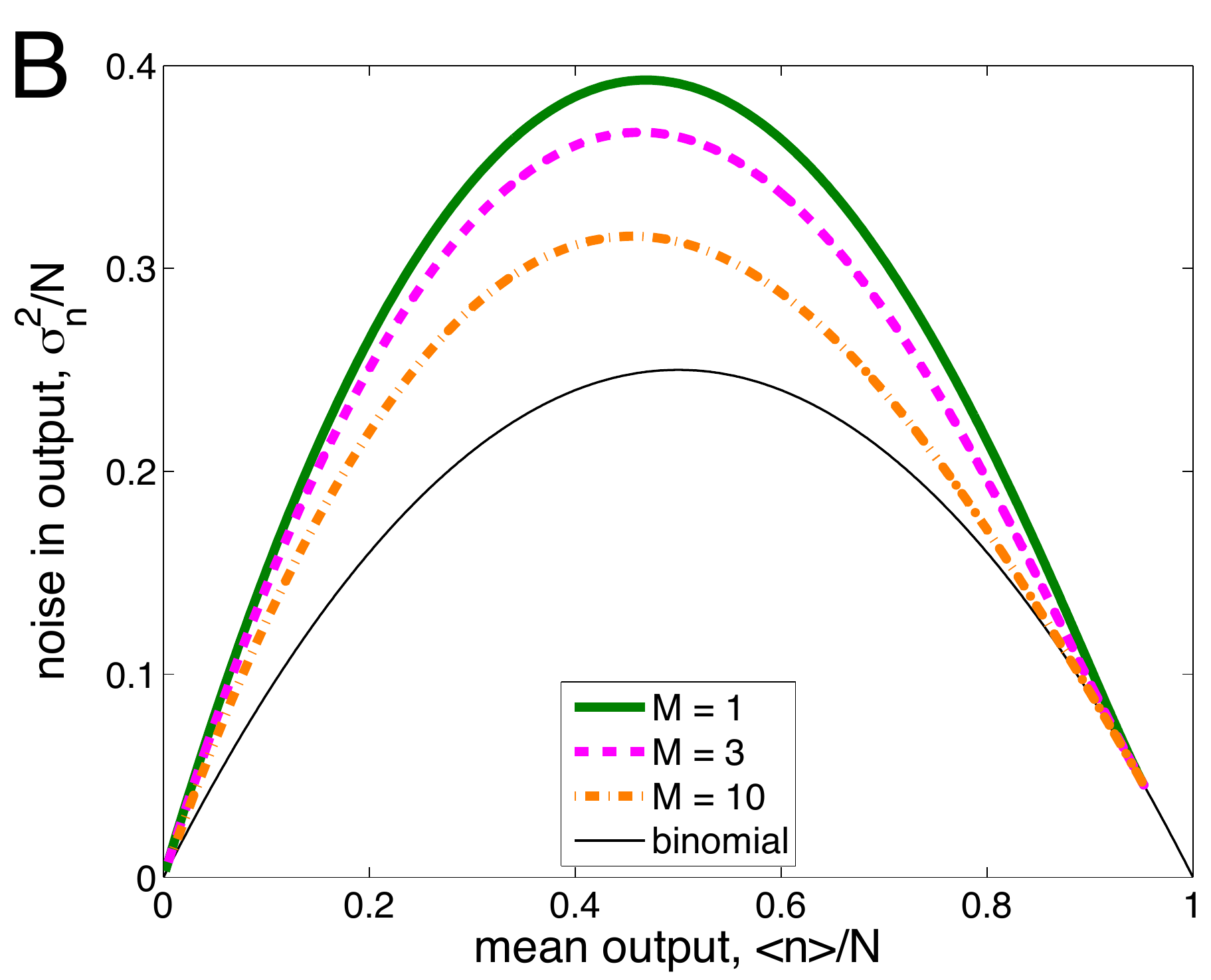}
	\end{center}
	\caption{	\label{fig:linearization}
		Reducing the number of input molecules linearizes the input-output response
		and increases the noise in the output.
Here $\pi=1$, $\beta=20$, and $\gamma=1$.
	}
\end{figure}

\vspace{-.2in}
		{\bf A} The output (the mean activity of $N=2$ $\cal{Y}$ molecules)
		vs.\ the input (the mean activity of $M$ $\cal{X}$ molecules) for several values of $M$.
		As $M$ is reduced the response becomes more linear,
		deviating more strongly from the mean-field response $\avg{n}/N = \beta q/(\beta q+1)$.
		Symbols show $20$ uniformly spaced values of $q$
		to highlight the effect of saturation on the state space.

		{\bf B} The noise vs.\ the mean for the output, shown
		for the same values of $M$.  As $M$ is reduced the noise increases
		for all values of the mean.
\clearpage

\ 
\begin{figure}[h]
	\begin{center}
		\includegraphics[width=0.75\textwidth]{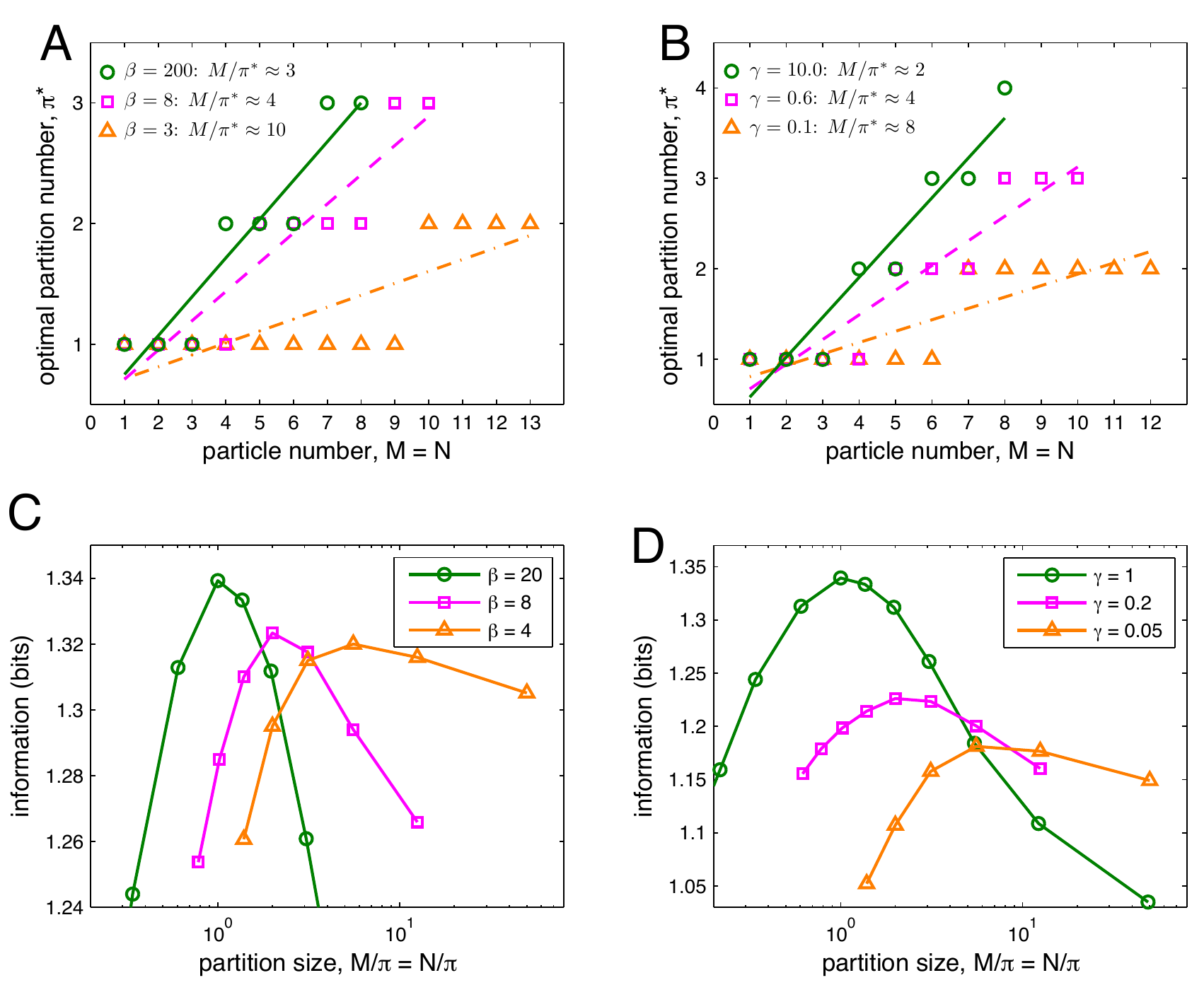}
	\end{center}
	\caption{	\label{fig:betagamma}
		The emergence of an optimal partition size is robust to parameter variations.$\qquad\qquad\qquad\qquad$
	}
\end{figure}

\vspace{-.2in}
		{\bf A, B} Results from the minimal system, described by the chemical master equation, as in Fig.\ 5B of the main text.  The information-optimal partition number $\pi^*$ is plotted as a function of
		molecule number $M=N$ for various values of $\beta$ (A) and $\gamma$ (B).
		Linear fits provide estimates of
		the optimal partition size $M/\pi^*$, as indicated in the legends.
		In A, $\gamma=1$; in B, $\beta=20$.

		{\bf C, D} Results from the lattice simulation, in which space is accounted for explicitly, as in Fig.\ 6C of the main text.  The information is plotted as a function of the partition size, directly revealing an optimum, for various values of $\beta$ (C) and $\gamma$ (D). Parameters are as in Fig.\ 6C: $M=N=49$, $p_{\rm hop}=0.001$, $\lambda=70$, and $p_D/p_r=1$.  In C, $\gamma=1$; in D, $\beta=20$.
		
As discussed in the main text, the optimum arises due to a tradeoff between two key effects of partitioning: on the one hand, partitioning removes correlations in the states of ${\cal Y}$ molecules, reducing noise; on the other hand, partitioning isolates molecules, reducing the maximal response.  The first effect favors few molecules per partition, while the second effect favors many molecules per partition.

As seen here in both the minimal system (A, B) and the simulated system (C, D), lowering $\beta$ or $\gamma$ increases the optimal number of molecules per partition.  This result has an intuitive explanation in terms of the above tradeoff: lowering either $\beta$ or $\gamma$ slows the rate of switching from the $Y$ to the $Y^*$ state, with respect to the timescale of ${\cal X}$ switching.  As a result, ${\cal Y}$ molecules are less sensitive to individual fluctuations in the state of ${\cal X}$ molecules.  The states of the ${\cal Y}$ molecules therefore exhibit weaker correlations, which in turn weakens the benefit that partitioning imparts in terms of the removal of these correlations.  The opposing effect of molecular isolation thus begins to dominate, pushing the optimum toward a larger number of molecules per partition.

\clearpage

\ \vspace{1in}
\begin{figure}[h]
	\begin{center}
		\includegraphics[width=0.6\textwidth]{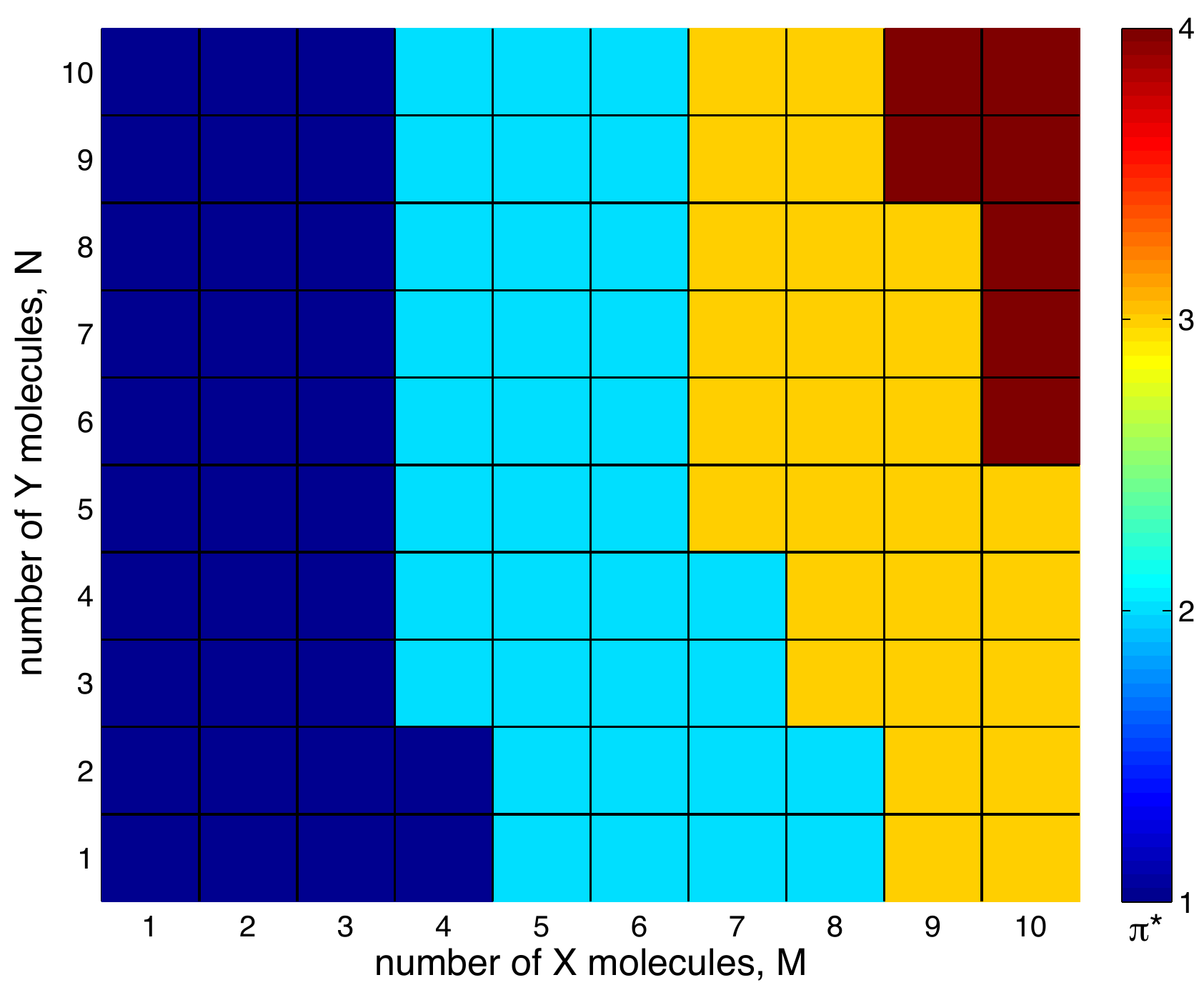}\\
	\end{center}
	\caption{	\label{fig:piMN}
		The optimal partition size has only a weak dependence
		on the number of output molecules.
		The information-optimal partition number $\pi^*$ is plotted as a function of
		the number of ${\cal X}$ molecules $M$ and the number of ${\cal Y}$ molecules $N$.
		The dependence of $\pi^*$ on $N$ is weak,
		such that the partition size $M/\pi^*\approx 3-4$ is roughly constant over the range of $N$ values.
	}
\end{figure}
\clearpage

\ \vspace{.5in}
\begin{figure}[h]
	\begin{center}
		\includegraphics[width=0.6\textwidth]{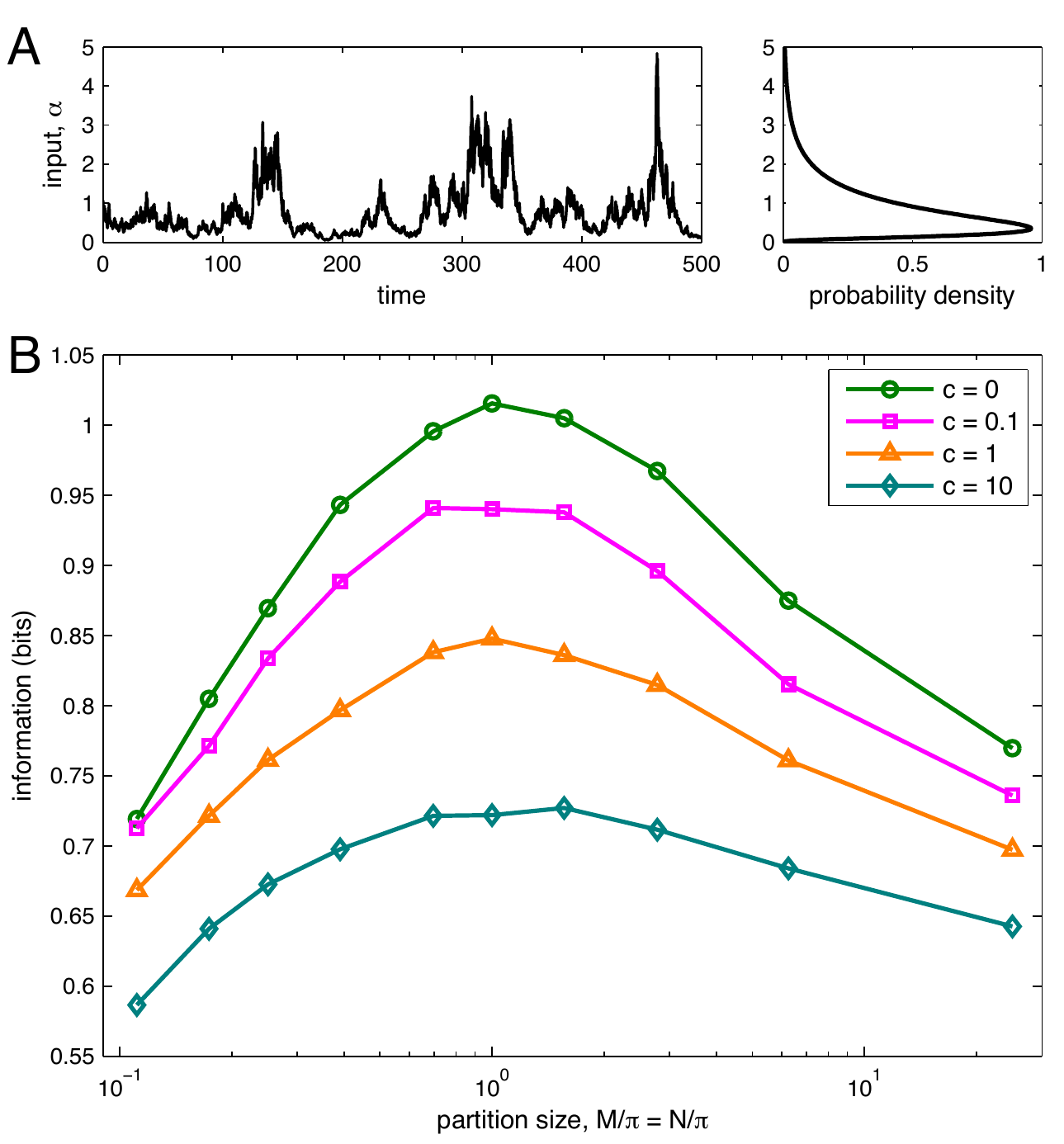}\\
	\end{center}
	\caption{	\label{fig:ex}
	The effects of partitioning are robust to extrinsic noise.$\qquad\qquad\qquad\qquad\qquad\qquad\qquad\qquad\qquad\qquad\qquad\qquad$
	}
\end{figure}

\vspace{-.2in}
	{\bf A} Simulations are performed with extrinsic noise introduced to the input parameter $\alpha$.  To keep $\alpha \ge 0$, the quantity $z \equiv \log \alpha$ is described by the simple mean-reverting Ornstein-Uhlenbeck process $dz = r(\mu-z)dt + \eta\sqrt{rdt}\xi$, where $\xi$ is a Gaussian random variable with mean $0$ and variance $1$; this results in a log-normal distribution for $\alpha$.  The quantity $1/r$ is the autocorrelation time, and the choices $\mu = \log[\bar{\alpha}^3/(\bar{\alpha}+c)]/2$ and $\eta = \sqrt{2\log(1+c/\bar{\alpha})}$ ensure that the mean of $\alpha$ is $\bar{\alpha}$ and that the variance of $\alpha$ scales with the mean via $\sigma_\alpha^2 = c\bar{\alpha}$.

	{\bf B} As the magnitude of the extrinsic noise (set by $c$) increases, the information $I[\bar{\alpha},n]$ decreases for all partition sizes, while the presence of an information-optimal partition size persists.

Here $M=N=25$, $\beta=20$, $\gamma=1$, $p_{\rm hop}=0.001$, the system is $\lambda = 50$ lattice spacings squared, the ratio of diffusion to reaction propensities is $p_D/p_r = 1$, and $r=1$ in units of the $X^*\to X$ reaction rate (which sets the timescale of switching). In A, $\bar{\alpha}=c=1$ and time is scaled by $1/r$.  In B, when $c=0$, the information transmission is lower than that in Fig.\ 6C of the main text because $M=N$ is lower.

\clearpage

\ \vspace{-.1in}
\begin{figure}[h]
	\begin{center}
		\includegraphics[width=0.9\textwidth]{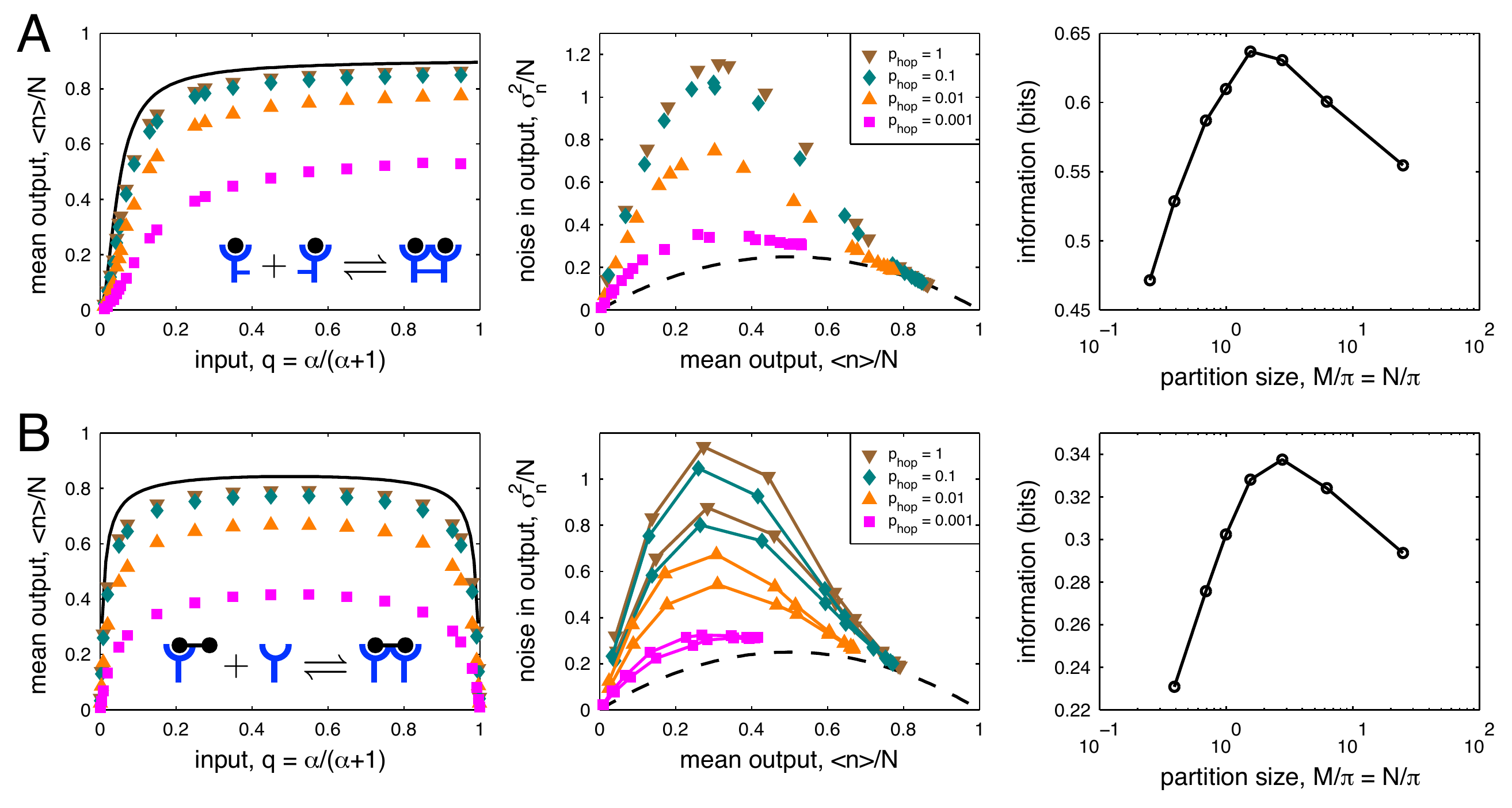}\\
	\end{center}
	\caption{	\label{fig:dimer}
	The effects of partitioning are robust to receptor dimerization.  Two dimerization schemes are simulated, which are paradigmatic for receptor tyrosine kinases, including EGF receptor \cite{Lemmon10}: {\bf A} Dimerization is receptor-mediated (left, inset), meaning two active receptors $X^*$ form a complex $C$, or {\bf B} dimerization is ligand-mediated (left, inset), meaning an active receptor $X^*$ and an inactive receptor $X$ form a complex $C$.  The latter scheme admits a ``dead-end'' state at ligand saturation, when all receptors are ligand-bound and no complexes can form, leading to a non-monotonic response curve (B, left), as observed e.g.\ for the Ret receptor \cite{Schlee06}.
	Both schemes are described by the reactions
$X\xrightleftharpoons[1]{\alpha}X^*$,
$C+Y\xrightarrow{\gamma\beta}C+Y^*$, and
$Y^*\xrightarrow{\gamma}Y$,
with dimer formation described by
$X^*+X^*\xrightleftharpoons[\chi]{\chi\epsilon}C$ in A, or
$X^*+X\xrightleftharpoons[\chi]{\chi\epsilon}C$ in B.
Here $M=N=25$, $\beta=20$, $\chi=\gamma=1$, the system is $\lambda = 50$ lattice spacings squared, and the ratio of diffusion to reaction propensities is $p_D/p_r = 1$.  In A, $\epsilon=20$; in B, $\epsilon=5$.
	}
\end{figure}

\vspace{-.2in}
{\bf Left} As in Fig.\ 6A of the main text, as the probability of crossing a diffusion barrier $p_{\rm hop}$ is decreased, the maximal value of the mean response decreases.  In A, the response also becomes more linear, but to less of a degree than in Fig.\ 6A of the main text.  Note that due to both finite diffusion and finite molecule number, even the unpartitioned response ($p_{\rm hop}=1$) deviates from the mean-field response (black solid line), which is given by $\avg{n}/N = \beta f/(1+\beta f)$, where $f$ is the fraction of ${\cal X}$ molecules in the dimer state; in A, $f = \epsilon g^2$ with $g \equiv \avg{m}/M = (\sqrt{1+8\epsilon q^2}-1)/(4\epsilon q)$, while in B, $f = \epsilon g(1-g)/(2\epsilon g+1)$ with $g \equiv \avg{m}/M = [\sqrt{1+8\epsilon q(1-q)}-1]/[4\epsilon (1-q)]$.  Here $\pi=25$.  Legends in middle panels apply to left panels as well.

{\bf Middle} As in Fig.\ 6B of the main text, as the probability of crossing a diffusion barrier $p_{\rm hop}$ is decreased, the output noise decreases.  Black dashed line shows the binomial noise limit $\sigma_n^2/N = (\avg{n}/N)(1-\avg{n}/N)$.  In B, lines connecting data points are provided to reveal that, as there are two values of $q$ that give the same mean $\avg{n}/N$ (left), the noise is higher for the smaller value of $q$.  Here $\pi=25$.

{\bf Right} As in Fig.\ 6C of the main text, the tradeoff between reliable signaling (reduced noise) and efficient signaling (maintaining a high maximal response) leads to an information-optimal partition size.  Here $p_{\rm hop}=0.001$.  Here, the information transmission is lower than that in Fig.\ 6C of the main text because $M=N$ is lower and additionally, in B, because of the non-monotonic mean response.

\clearpage

\ \vspace{1in}
\begin{figure}[h]
	\begin{center}
		\includegraphics[width=0.9\textwidth]{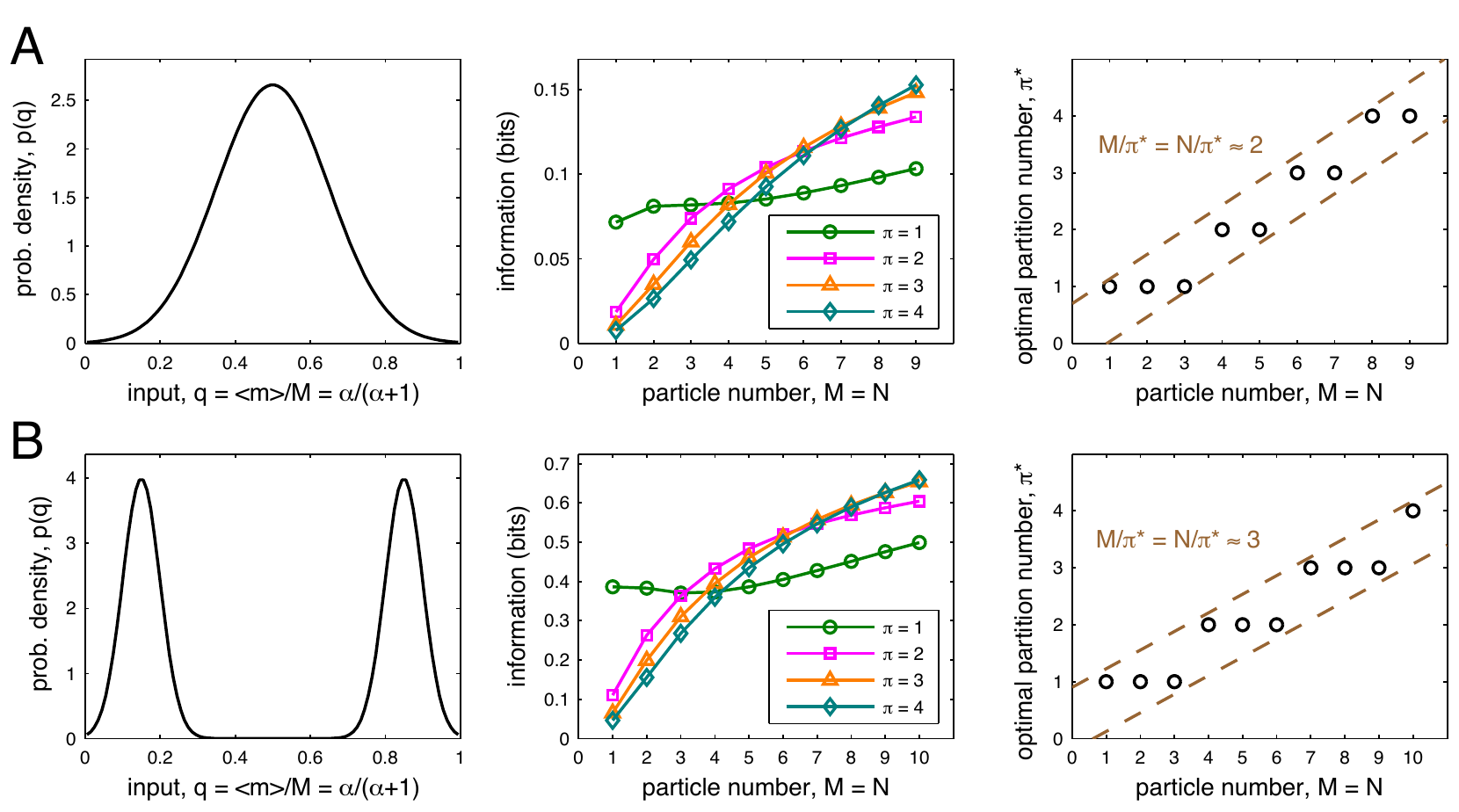}
	\end{center}
	\caption{	\label{fig:input}
		The effects of partitioning are robust to the shape of the input distribution.  As in Fig.\ 5 of the main text, which takes a uniform input distribution $p(q)$, an information-optimal partition size $M/\pi^* = N/\pi^*$ persists with an input distribution that is {\bf A} unimodal or {\bf B} bimodal.
	}
\end{figure}
\clearpage

\ \vspace{1in}
\begin{figure}[h]
	\begin{center}
		\includegraphics[width=0.6\textwidth]{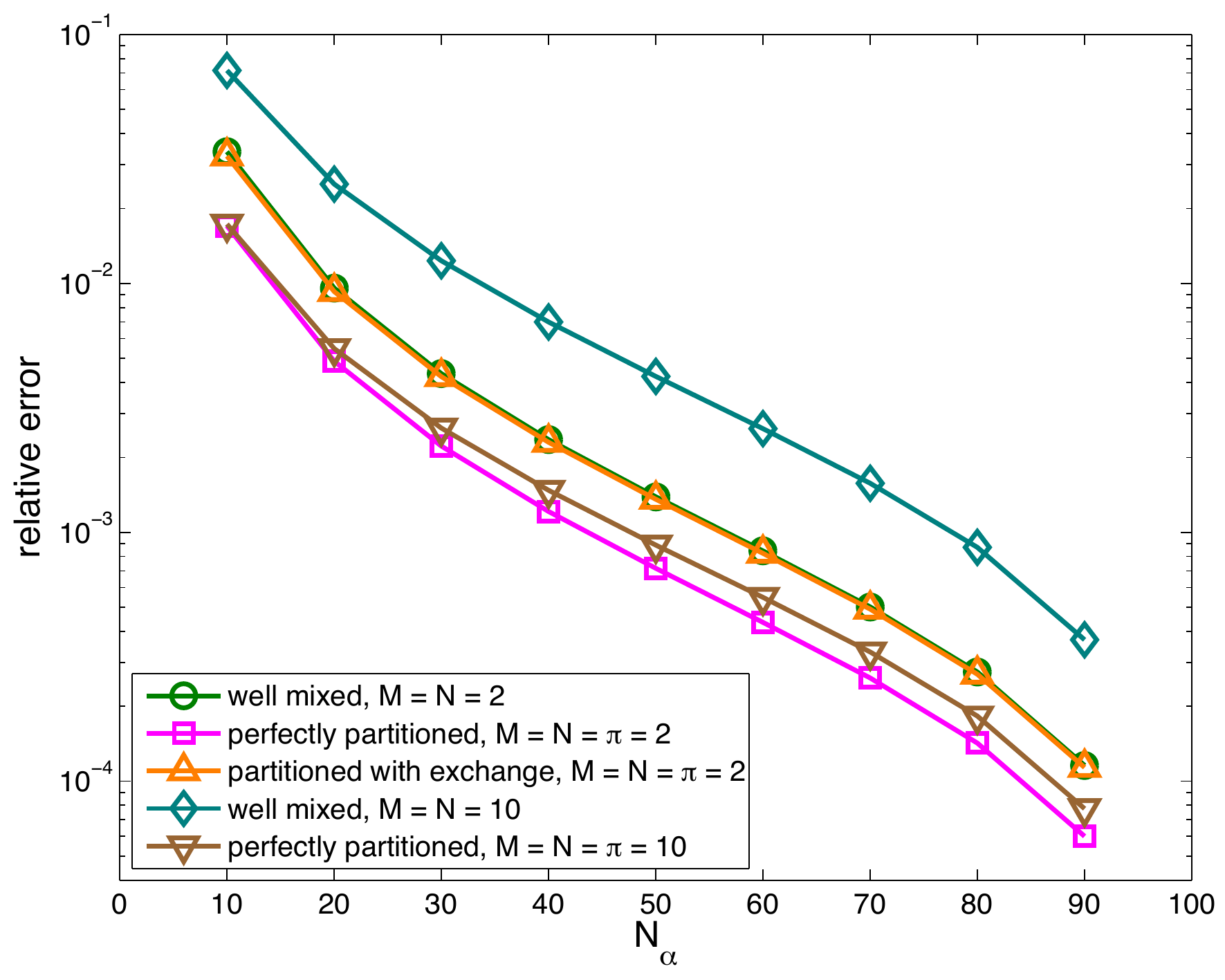}
	\end{center}
	\caption{	\label{fig:Nalpha}
		Computation of the mutual information converges as the input is more finely discretized.
		The relative error $\abs{I-I_0}/I_0$, where $I_0$ is the information at $N_\alpha = 100$,
		is plotted against the number $N_\alpha$ of values of $\alpha$
		[uniformly spaced in $q = \alpha/(\alpha+1)$] used in the computation.
		Five conditions are tested, as indicated in the legend.
		It is seen that the relative error falls below $\sim$$1\%$ in all conditions
		for $N_\alpha \gtrsim 30$.
	}
\end{figure}

\end{document}